\documentclass[10pt,a4paper,twocolumn]{article}
\usepackage{derivative}

\usepackage[scaled]{helvet} 
\usepackage[T1]{fontenc} 

\usepackage{graphicx}
\usepackage[english]{babel}
\usepackage{dcolumn}
\usepackage{bm}
\usepackage{url}
\usepackage[numbers,square,sort&compress]{natbib}
\usepackage{amsmath}
\usepackage[usenames,dvipsnames]{xcolor}
\usepackage{soul}
\usepackage{subcaption,graphicx}
\usepackage{tabularray} 

\input{pisika1.dat}

\usepackage{float}   

\begin{document}

\providecommand{\ShortAuthorList}[0]{A.~M.~Surname, B.~D.~Suffix Jr., C.~G. Suffix III} 
\title{Atomistic Simulations of Wetting Dynamics of Water Nanodroplets on Nanotextured Titanium: Implications for Medical Implants}
\author[1]{Ilemona S. Omeje\thanks{Corresponding author: ilemona.sunday.omeje@univ-st-etienne.fr}}
\author[2]{Djafar Iabbaden}
\author[3]{Patrick Ganster}
\author[1]{Tatiana E. Itina}
\affil[1]{Université Jean Monnet, CNRS, Institut d’Optique Graduate School, Laboratoire Hubert Curien, Saint-Etienne, France} 
\affil[2]{LEM3, CNRS – Université de Lorraine – Arts et Métiers ParisTech, 7 rue Félix Savart, 57070 Metz, France}
\affil[3]{Mines Saint-Étienne, Univ. Lyon, CNRS, UMR 5307 LGF, Centre SMS, F - 42023
Saint-Étienne, France}


\date{}

\begin{abstract}
\noindent

The development of high-performance biomedical implants requires a deep understanding of the molecular interactions between water molecules and titanium (Ti) surfaces. In this study, fully atomistic molecular dynamics simulations were used to study the static and dynamic wetting behavior of water nanodroplets on both flat and femtosecond laser-induced nanotextured Ti surfaces. Our findings reveal a clear transition from Wenzel to Cassie-Baxter wetting states as surface roughness increases, significantly affecting droplet spreading. We also observe the damping of nanodroplet vibrations and a roughness-dependent shift toward hydrophobicity, driven by stronger atomic interactions between water molecules and surface atoms. Furthermore, the interaction energy between water droplets and nano-textured Ti surfaces decreases with increasing roughness, reinforcing the observed changes in wettability. The discrepancies observed between classical wetting models and nanoscale behavior emphasize the limitations of current theoretical approaches and the importance of developing more advanced models. This study provides valuable insights into optimizing Ti surface properties for improved implant performance through controlled wettability.

\keywords{titanium, wetting, contact angle, femtosecond laser, roughness, molecular dynamics, water, nanodroplet}

\DOI{} 
\end{abstract}

\maketitle
\thispagestyle{titlestyle}

\section{Introduction}\label{sec:intro}

Titanium (Ti) and its alloys hold immense promise, particularly in biomedical applications, due to their exceptional biocompatibility, high strength, and fracture toughness \citep{elias2008biomedical}. These properties make them ideal candidates for dental and orthopedic implants. A well-established strategy for improving the integration of Ti implants with bone involves the introduction of appropriate surface roughness \citep{kligman2021impact}. Surface texturing has also been shown to improve antibacterial properties \citep{lutey2018towards}. Among the various surface treatment methods, femtosecond laser processing stands out due to its high processing resolution, minimal heat-affected zone \citep{vorobyev2013direct}, and the ability to fabricate complex 2D and 3D structures \citep{yong2018hall, sugioka2014ultrafast}.

The wettability of laser-treated implant surface plays a pivotal role in influencing its interaction with surrounding tissues, promoting cell attachment, and preventing bacterial growth \citep{kligman2021impact, gittens2014review}. A comprehensive understanding of the wetting properties of Ti surfaces is important. While several models have been developed to account for surface topography and heterogeneities \citep{quere2008wetting}, the experimentally observed apparent contact angle (CA) on femtosecond laser-treated surfaces often deviates from the predictions of Young's model \cite{wu2009superhydrophobic, cunha2013wetting, raimbault2016effects, dou2020influence}. This discrepancy is typically explained using the Wenzel and Cassie-Baxter equations, which consider surface irregularities \cite{wenzel1936resistance, cassie1944wettability}. However, these models are primarily applicable to homogeneous and isotropic surfaces with large droplet-to-roughness scale ratios, and may not fully capture the variations of heterogeneous micro/nanostructures created by femtosecond lasers \cite{bizi2011modifications}. On the other hand, studies have shown that micro/nanostructures fabricated using femtosecond lasers exhibit dynamic wetting properties, where the contact angle of a liquid changes over time due to advancing and receding contact angles, influenced by surface topography and energy \citep{yong2018hall, yong2014simple, giannuzzi2019short}.

The challenges associated with nano-surface processing and experimentation at the nanoscale have significantly limited our understanding of nanodroplet interactions with textured surfaces. To overcome these limitations, Molecular Dynamics (MD) simulations have emerged as a powerful tool for investigating the wetting properties of nanodroplets on nanometer-scale surface reliefs. This approach holds significant potential for various applications, particularly in understanding how nanostructures created by femtosecond laser texturing affect the attachment of cells to medical implants \cite{cunha2016femtosecond, dumas2012multiscale, sirdeshmukh2021laser}. By optimizing these surface textures, femtosecond laser technology can promote specific cellular responses, improving the integration of artificial cells with medical implants and advancing the overall success of implant procedures \cite{shivakoti2021laser, mukherjee2015enhancing}.

Previous research has extensively studied the wetting behavior of titanium dioxide (TiO$_2$) using MD simulations. For example, \citet{park2009temperature} used MD simulations to show that the wettability of a TiO$_2$ surface increases with rising temperature due to a decrease in hydrogen bonding and surface tension. Similarly, \citet{ohler2009molecular} proposed a new model for the CA of water on TiO$_2$ surfaces using MD simulations. They argue that the macroscopic CA is not determined by the direct interaction of bulk water with the TiO$_2$ surface but rather by an interface between a thin, ordered layer of water molecules on the surface and the bulk water droplet. 
 
Recently, \citet{zhu2023surface} investigated the wettability of three different phases of TiO$_2$ thin films: anatase, brookite, and rutile. The study found that the wettability of the TiO$_2$ surface depends on the crystal phase, with anatase being the most wettable (CA of 43.7$^\circ$) and rutile the least wettable (CA of 73.9$^\circ$). The authors attribute this difference to variations in the structural connection and arrangement of surface microtopography among the three phases. The study also found that increasing the roughness of the anatase surface decreases its wettability, further highlighting the role of surface morphology in determining wettability. However, despite the wealth of research on TiO$_2$, the wettability of pure Ti remains largely unexplored.

In this study, we address this gap by employing MD simulations to investigate the wettability of water nanodroplets on Ti, both on flat and nano-textured surfaces, mimicking the rippled metasurfaces commonly produced by femtosecond laser processing with structure sizes of a few tens of nanometers. This research provides new atomic-level insights into how water interacts with Ti, particularly under varying surface roughness. Our study reveals the influence of nanostructuring on the wettability of Ti, which is essential for optimizing surface properties for medical implants. Additionally, by examining both static and dynamic wetting phenomena, we provide a comprehensive understanding of the wetting mechanisms, thus offering data that can bridge the gap between classical models and experimental observations.

\begin{table}[h]
  \centering
  \begin{tblr}{
      width = \linewidth,
      colspec = {|Q[1,c,m]|Q[1,c,m]|Q[1,c,m]|Q[1,c,m]|},
      hlines,
      vlines,
      row{1} = {font=\bfseries},
      row{odd} = {gray9},
    }
    \hline
     Pair & $\sigma$ (nm) & $\varepsilon$ (kcal/mol) & Charge (e) \\
    
    $O - O$ & 0.3166 & 0.1554 &  -0.8476\\
    $H -  H$ & 0.0 & 0.0 & +0.4238\\
    $Ti -  Ti$ & 0.2958 & 0.8599 & 0\\
    $Ti -  O$ & 0.3062 & 0.3656 & 0\\
    \hline
  \end{tblr}
  \caption{SPC/E water and Ti force field interaction parameters values.}
  \label{tab:SPCE_Parameter}
\end{table}

\section{Simulations model and details}\label{sec:2}
The water nanodroplet on the Ti surface is modeled using the Extended Simple Point Charge (SPC/E) model to represent liquid water \cite{berendsen1987missing}. The SPC/E model is a three-site rigid model with point charges on Oxygen (O) and Hydrogen (H) (Table \ref{tab:SPCE_Parameter}). This model have been extensively used in wetting studies \cite{chen2020mechanism}. In this model, the interactions between water molecules are described using a combination of short-range Lennard-Jones (LJ) and long-range electrostatic potential energy, as given by the following equation:

\begin{equation}
\label{eq:law1}
\begin{aligned}
U(r_{ij}) = U_{non-bond}(r_{ij}) + U_{e}(r_{ij}) \\ = 4\varepsilon_{ij}\left[\left(\frac{\sigma_{ij}}{r_{ij}}\right)^{12} - \left(\frac{\sigma_{ij}}{r_{ij}}\right)^6\right] \\
+ \frac{1}{4\pi\epsilon}\sum_{i=1}^{3}\sum_{j=1}^{3}\frac{q_i q_j}{r_{ij}}
\end{aligned}
\end{equation}

The short-range LJ potential ($U_{non-bond}(r_{ij})$) represents the Van der Waal force between two non-bonding atoms. As shown in Equ. \ref{eq:law1}, it is defined by a conventional 12-6 function such that $r_{ij}$ is the distance between the two atoms i and j, $\sigma_{ij}$ is the distance at which the interaction energy between two atoms is minimal, and $\varepsilon_{ij}$ is the depth of the potential well. In the Coulombic term ($U_{e}(r_{ij})$), $\epsilon$ represents the permittivity in vacuum, $q_i$ and $q_j$ denote the charges of the interaction sites. The OH bond length is 1 \r{A} and the angle between H-O-H is fixed at 109.47$^\circ$. To make the water molecule more rigid during the MD simulation, the SHAKE algorithm was used to constrain bond length and angle \cite{ryckaert1977numerical}. Long-range electrostatic interactions were computed using the Particle-Particle Particle-Mesh (PPPM) method \cite{hockney2021computer}. The PPPM method is a distinct method for efficiently computing long-range electrostatic interactions in periodic systems \citep{darden1993particle}. It offers a complementary approach to the Ewald method, often demonstrating superior computational performance for large systems \citep{barisik2013wetting}. Non-bonding interactions, including both Van der Waals and electrostatic forces, were truncated at a cut-off distance of 12 \r{A}. The interactions between Ti and O atoms were modeled using the 12-6 LJ potential, with parameters determined by the Lorentz-Berthelot mixing rule \cite{allen1987computer}, as shown in Equ. \ref{eq:law2}.

\begin{equation}
\label{eq:law2}
\sigma_{ls}=\frac{\sigma_i + \sigma_j}{2}; \quad \varepsilon_{ls}=\sqrt{\varepsilon_i \cdot \varepsilon_j}
\end{equation}

It should be noted that the interactions between Ti and H atoms are neglected in this study. This simplification based on the omission of LJ parameters for H atoms in the SPC/E water model is a well-established practice supported by extensive research and literature. The primary reasons for this simplification are computational efficiency and the focus to accurately capture the critical electrostatic interactions that determine the behavior of the water molecules \citep{allen2017computer, guillot2002reappraisal}. Additionally, because the bonds in the SPC/E model of water are rigid, the H atoms are restricted from freely rotating and forming independent interactions with Ti \citep{berendsen1987missing}. Therefore, the H atoms are primarily involved in H bonding with neighboring water molecules \citep{allen2017computer}, further reducing their availability for direct interaction with Ti. It is worth mentioning that neglecting H-substrate interaction may underestimate the influence of wetting behavior on highly reactive (laser-treated) surfaces that can readily form H bonds with the H atoms of water. The LJ interaction potential parameters for Ti and water is found in Table \ref{tab:SPCE_Parameter}.

\begin{table}[h]
  \centering
  \begin{tblr}{
      width = \linewidth,
      colspec = {|Q[1,c,m]|Q[1,c,m]|Q[1,c,m]|Q[1,c,m]|Q[1,c,m]|},
      hlines,
      vlines,
      row{1} = {font=\bfseries},
      row{odd} = {gray9},
    }
    \hline
     Surface (Surf) & h (nm) & w (nm) & s (nm) & p (nm) \\
    Surf-1 & 0 & 0 & 0 & 0 \\
    Surf-2 & 0.28 & 0.911 & 1.109 & 2.02 \\
    Surf-3 & 0.28 & 0.509 & 0.511 & 1.02 \\
    Surf-4 & 0.78 & 1.011 & 1.009 & 2.02 \\ 
    Surf-5 & 0.78 & 0.509 & 0.511 & 1.02 \\
    \hline
  \end{tblr}
  \caption{ Values of the adjustable dimensional parameters such as height (h), width (w), spacing (s) and periodicity (p) of fence-like Ti nanostructures.}
  \label{tab:roughness_dimension}
\end{table}

\begin{figure}[h]
\centering
\includegraphics[width=1\linewidth]{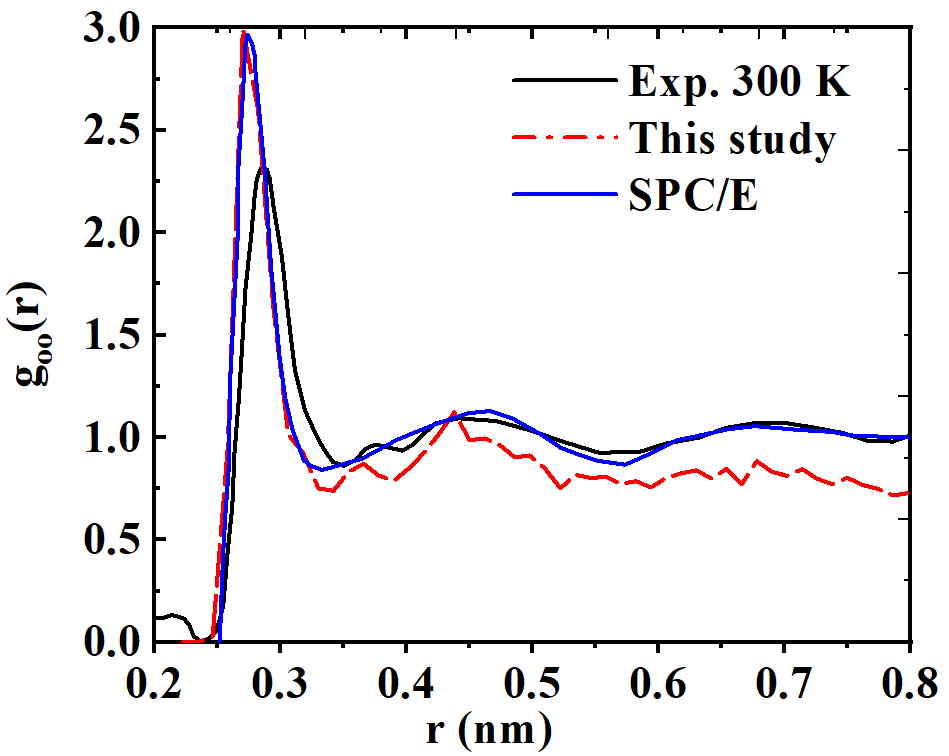}
\caption{Radial distribution functions (g(r)) of oxygen-oxygen distances for real water \citep{narten1971liquid}, this MD study and SPC/E water model \citep{berendsen1987missing} at 300 K.} 
\label{fig:RDF}
\end{figure}

\begin{figure}[h]
\centering
\includegraphics[width=1\linewidth]{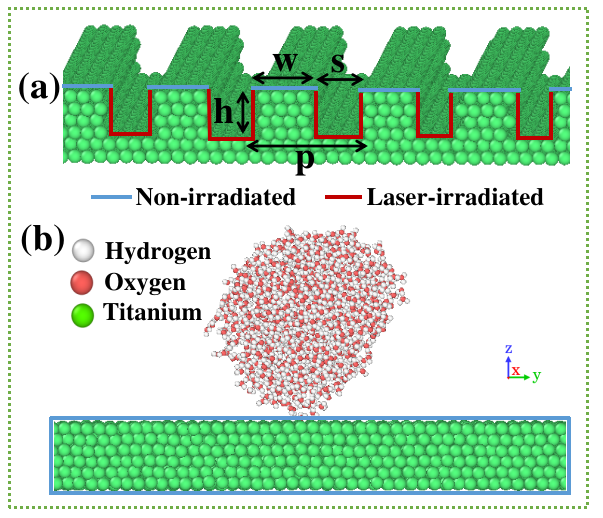}
\caption{(a) A fence-like nanostructure on Ti surface with heights ($h$), widths ($w$), spacing ($s$), and the periodicity ($p$). The blue and red markers indicate non-irradiated and laser-irradiated regions, respectively, mimicking the rippled metasurfaces commonly produced by femtosecond laser processing. (b) Initial atomistic configuration of water nanodroplet on the flat Ti surface.}
\label{fig:dropTi_rough}
\end{figure}

In addition, Ti-Ti interactions were calculated using the LJ potential. We initially tried LJ potentials from other sources \citep{park2009temperature, ohler2009molecular, zhu2023surface,bandura2003derivation} including the Universal Force Field (UFF) \citep{rappe1992uff}. The LJ potential parameters for Ti interactions used in this study are modified from the UFF in such a way that they can replicate the physical properties of Ti sufficient for Ti-water interactions that best aligns with existing calculations of water nanodroplets on materials \citep{chen2020mechanism}. The parameters have been validated against Density Functional Theory (DFT) calculations \citep{wang2021diamond, ikehata2004first} and experimental data \citep{hua2017composite, fisher1964single}. These parameters accurately reproduce the lattice parameters and independent elastic constants of Ti.

The nanodroplet corresponds to a sphere with a radius of 3.0 $nm$ and contains $3009$ atoms. Previous studies have consistently demonstrated that droplets comprising 2000 atoms (or more) consistently produce reliable results \citep{song2013molecular}. Temperature equilibration of the nanodroplet and the Ti substrate was performed separately. The nanodroplet was equilibrated using a time step of 1 $fs$ at 300 $K$ for approximately 1 $ns$ in the NVT ensemble. This duration was sufficient to achieve a stable nanodroplet configuration. To validate this configuration, the radial distribution functions (g(r)) obtained from our simulations were compared with the experimental data \citep{narten1971liquid} and the SPC/E model \citep{berendsen1987missing}, as shown in Fig. \ref{fig:RDF}. The comparison indicates good agreement between our study, the experimental data, and the SPC/E model with slight deviations observed below the peaks of g(r). This agreement demonstrates that the SPC/E water model used in our MD simulations is appropriate and reliable.

\begin{figure*}[th!]
\centering
\includegraphics[width=1\linewidth]{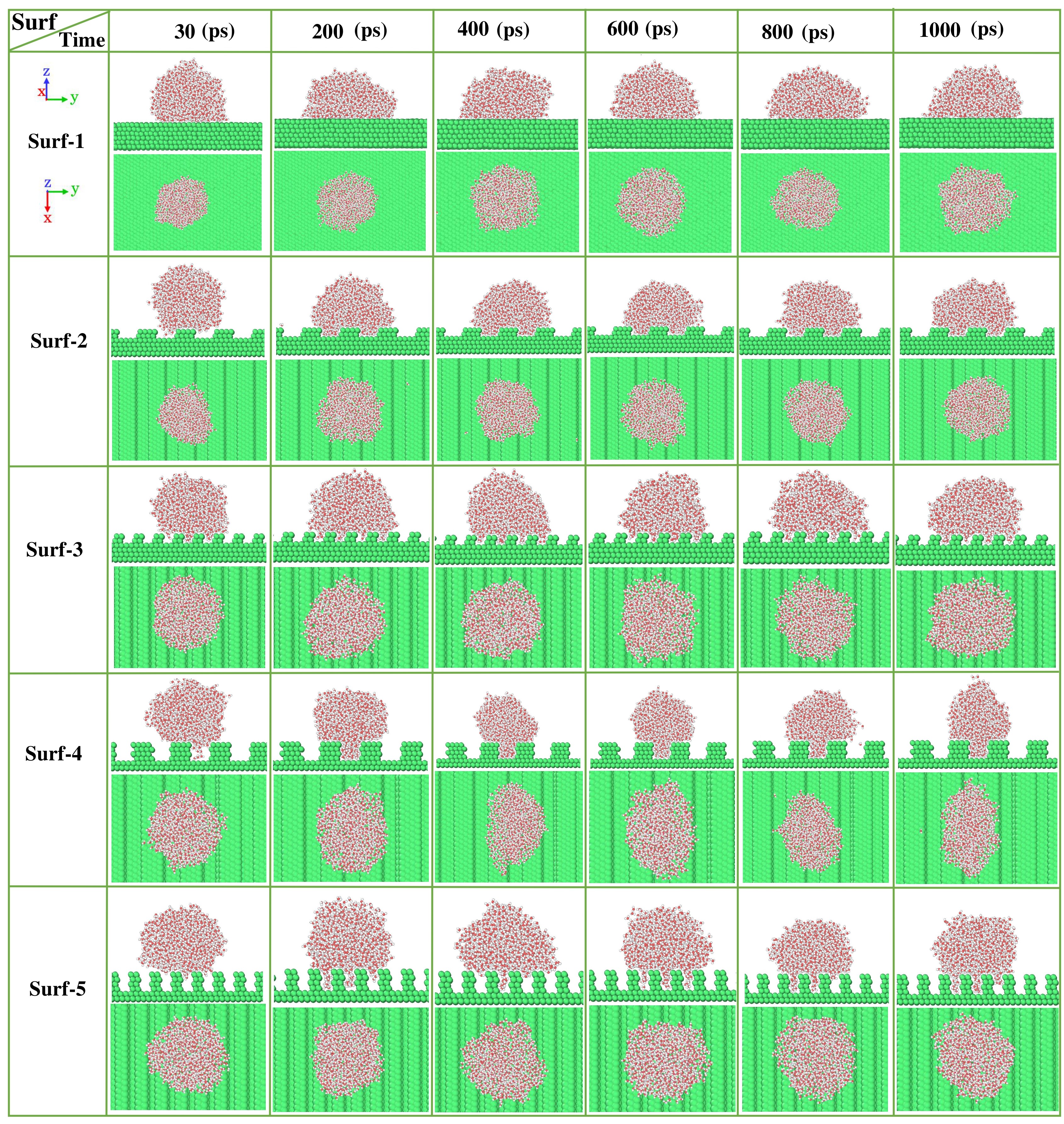}
\caption{Snapshots of the side and front views of the spreading process of a water nanodroplet on Ti surfaces with different roughness scales as a function of time. The roughness parameters used in the simulations are shown in Table 2: h (height), w (width), s (spacing) and period (p) vary across surfaces, with h ranging from 0.28 nm to 0.78 nm, w from 0.509 nm to 1.011 nm, s from 0.511 nm to 1.109 nm, and p from 0 to 2.02 nm respectively.}
\label{fig:Dy1}
\end{figure*}

The flat Ti substrate corresponds to a block of dimension $14.5$ $\times$ $19.5$ $\times$ $1.4$ nm\textsuperscript{3} containing $22344$ atoms arranged in a Hexagonal Close-Packed (HCP) structure along [0-110] [10-10] [0001] directions. Subsequently, a nanometric pattern similar to laser-textured ones, having heights ($h$), widths ($w$), spacing ($s$), and a periodicity ($p$) as depicted in Fig. \ref{fig:dropTi_rough} (a), was introduced on Ti surfaces \citep{predictwetting, niu2014static, zhao2021molecular}. In our previous studies, it has been established that the structural parameters $h, w, s,$ and $p$ are adjustable based on the topography profile of the laser-patterned surfaces \citep{omeje2022numerical, predictwetting}. The adjustable parameters used in the wetting studies are summarised in Table \ref{tab:roughness_dimension}. For the sack of clarity and to facilitate the adjustment of these parameters, we use the roughness factor ($r$) and fractional area ($f$) to quantify them, as shown in Equ. \ref{eq:law3} below.

\begin{equation}
\label{eq:law3}
r = 1 + \frac{2h}{p}; p = w+s; f = \frac{w}{p}
\end{equation}

The bulk  Ti substrate was equilibrated using the NVT ensemble at 300 $K$ for 1 $ns$ with a time step of 1 $fs$. In addition, the equilibrated spherical nanodroplet was positioned on the equilibrated Ti surface, as shown in Fig. \ref{fig:dropTi_rough} (b). In this study, the flat and rough Ti substrates were completely frozen during the MD simulation. The fixed substrate in MD simulations, particularly in wetting studies, is a well-documented practice aimed at a reduced computational cost \citep{allen2017computer}, maintaining numerical stability \citep{frenkel2023understanding} and ensuring an accurate representation of the physical system \citep{yaghoubi2018molecular, koishi2009coexistence}. Interestingly, simulations with a fixed surface compared to a flexible surface have shown a negligible effect on the CA \cite{chen2020mechanism}. 

In addition to the simulation set-up, the size of the simulation box was set to $14.5$ $\times$ $19.5$ $\times$ $15.4$ nm\textsuperscript{3}.  Periodic boundary conditions were applied in the x, y, and z directions during the simulation of the water nanodroplet on the Ti surfaces. The entire simulation was carried out for 1 ns in the NVT ensemble with a time step of 1 fs. Temperature control was achieved using the Nosé-Hoover thermostat to maintain a constant temperature of 300 K. The Velocity Verlet algorithm was used for numerical integration of the equations of motion. MD calculations were performed using the Large-scale Atomic/Molecular Massively Parallel Simulator (LAMMPS) software \cite{plimpton1995fast}. Visualization and analysis of the water nanodroplet on Ti surfaces were conducted using OVITO software \cite{stukowski2009visualization}.

\section{Results and discussion} \label{sec:3} 

In this section, we present the results of MD simulations of the model described above to investigate the wettability properties of both flat and nanorough Ti surfaces. Initially, we focused on the dynamic wetting behavior on both Ti surfaces by analyzing the spreading radius and dynamic CA of a nanodroplet. These analyses enabled us to determine the equilibrium state of the nanodroplet on the Ti surface. Subsequently, we studied the wettability of nanodroplets on rough Ti surfaces by evaluating the static CA and the influence of surface nanoroughness on the wetting properties. The results obtained were compared to existing classical wetting models such as the Wenzel and the Cassie-Baxter model.

\subsection{ Dynamic wetting behavior of Ti surfaces}

To understand the fundamental process of nanodroplet spreading on the flat and nanometric Ti surfaces, the dynamics of the nanodroplet on the corresponding surfaces are examined.  Thus, Fig. \ref{fig:Dy1} shows the snapshots of the side and front views of the spreading process of a water nanodroplet on Ti surfaces with different roughness scales as a function of time. The droplet on the flat surface (Surf-1) is initially spherical at 0 ps, indicating minimal spreading. Between 30 and 50 ps, the droplet starts to spread gradually, with a noticeable decrease in height. From 200 to 1000 ps, further spreading occurs, and the droplet becomes more stable with a large base and minimal height.

\begin{figure}[h]
\centering
\includegraphics[width=1\linewidth]{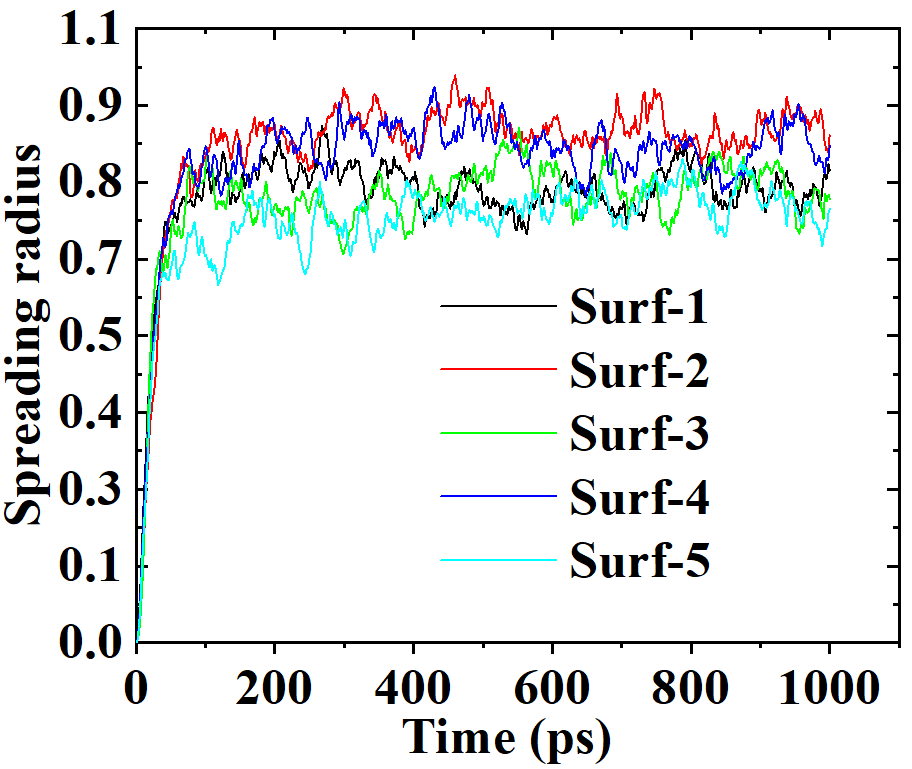}
\caption{Normalized spreading radius of water nanodroplet on Ti surfaces with different roughness.}
\label{fig:Sp}
\end{figure}

In the case of Surf-2, the droplet spreads and begins to interact with the rough surface by 30 ps. As time progresses, between 200 and 1000 ps, the droplet flattens out, forming a wide base that settles into the roughness grooves. This indicates a substantial degree of spreading. In Surf-3, the droplet spreading is restricted compared to Surf-2 and Surf-4. The narrow width and height of the grooves beneath the droplet in Surf-3 limit the extent to which the droplet can spread. The droplet conforms to these grooves, resulting in less lateral spreading compared to surfaces with larger grooves like Surf-2. This behavior is due to the greater groove density in Surf-3, as indicated by the equal distribution of w and s. 

In Surf-4, the droplet spreads more extensively than in previous cases. Initially, from 30 to 200 ps, the droplet expands equally in both the x and y directions, filling one of the roughness grooves. This notable difference in spreading behavior can be attributed to the larger dimensions of the grooves in Surf-4, specifically the h, w, and s, as shown in Table \ref{tab:roughness_dimension}. While the groove width and spacing in Surf-4 are similar to those in Surf-2, the grooves are deeper (greater h) in Surf-4. By 1000 ps, the droplet's base shrinks along the y-direction leading to a noticeable increase in the droplet's height along this axis while it continues to spread along the x-direction. However, for Surf-5, the droplet partially retracts, forming a ball-like shape with air pockets trapped beneath its surface. This altered spreading pattern is due to the significantly deeper grooves (higher h) in Surf-5, coupled with narrower w and s compared to Surf-2, 3 and 4. 

It is therefore important to state that the degree of roughness and the specific dimensions of the surface grooves (h, w, s and p) critically determine the wetting process. Surfaces with larger grooves and spacing (like Surf-2 and Surf-4) tend to support more extensive spreading, while those with narrower grooves and higher roughness (like Surf-3 and Surf-5) limit spreading.

\begin{figure}[h]
\centering
\includegraphics[width=1\linewidth]{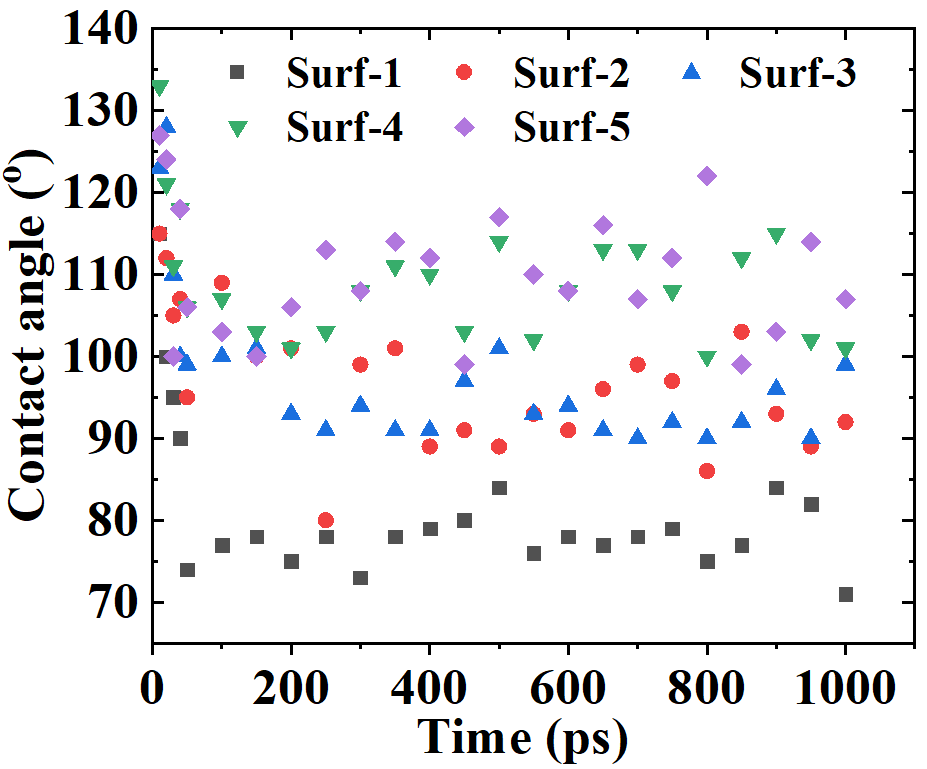}
\caption{Dynamic contact angles of water nanodroplet at different times on Ti surfaces with different roughness scales.}
\label{fig:Dy}
\end{figure}

\begin{figure*}[h]
\centering
\includegraphics[width=1\linewidth]{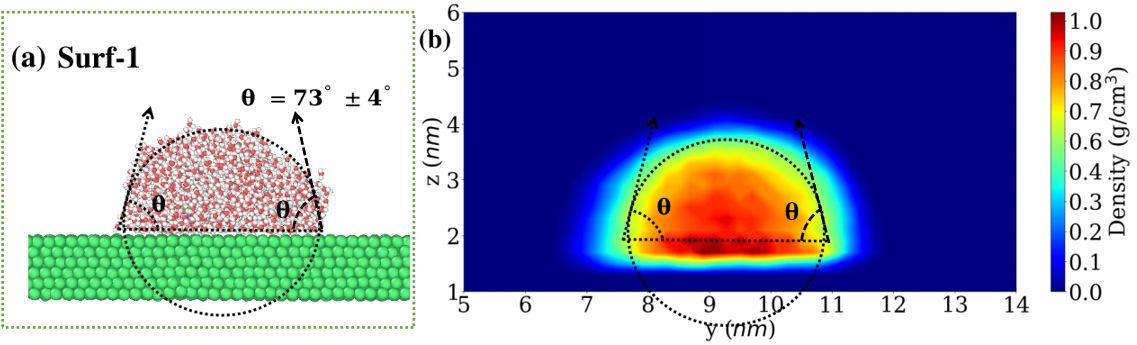}
\caption{Measurement of static water contact angle: (a) snapshot of a water nanodroplet and (b) its density profile on a flat Ti surface with an indication of the water contact angle. In the density profile color bar, 1 corresponds to the liquid and 0 corresponds to the vapor.} 
\label{fig:density}
\end{figure*}

Fig. \ref{fig:Sp} presents the normalized spreading radius as a function of time for water nanodroplets on Ti surfaces with varying roughness scales. Initially, the spreading radius increases rapidly, indicating that the droplets begin to spread quickly upon contact with the surface. During the first 50 ps, there is a significant increase in the spreading radius for all surfaces. Beyond 50 ps, the spreading radius begins to approach a steady state.

It is observed that Surf-5 has the lowest spreading radius, followed by Surf-3. These observations validate the earlier claim about the similar roughness features of the two surfaces. Surf-2 has the highest spreading radius, followed by Surf-4. This further validates the results mentioned above concerning the two surfaces. The spreading radius of the flat surface (Surf-1) is lower than that of Surf-2 and 4 but higher than that of Surf-3 and 5. While the increased roughness in Surf-3 and 5 limits spreading, the spreading radius of Surf-2 and 4 is higher than that of Surf-1 because, as the droplet stabilizes, the droplet base on Surf-2 and 4 is larger compared to Surf-1.

Fig. \ref{fig:Dy} illustrates the dynamic CA of water nanodroplets as a function of time for Ti surfaces with varying roughness levels. The dynamic CA was determined by taking snapshots of the nanodroplets at regular intervals and measuring the advancing and receding CAs specifically in the Y-Z direction neglecting X-Z direction. The difference between the advancing and receding contact angles is represented by the error bars. Initially, the droplets exhibit minimal spreading on all surfaces, with the CA at 30 ps ranging from approximately $115^\circ$ for the flat surface (Surf-1) to around $133^\circ$ for Surf-4. The CA decreases over time, signifying the spreading of the droplets, with the most rapid decrease occurring within the initial 50 ps. This decrease is due to the dynamic rearrangement of water molecules at the liquid-Ti interface, driven by intermolecular forces and the electrostatic interaction between the oxygen atom of the water molecule and the Ti atoms. This molecular rearrangement lowers the interfacial energy, resulting in a decrease in the CA, which leads to a more stable and spread-out droplet configuration. Notably, Surf-5, with the highest roughness, consistently exhibits a higher CA compared to the other surfaces, indicating that increased roughness hinders wetting and promotes hydrophobicity due to the trapping of air pockets beneath the droplet.

 \begin{figure*}[h]
  \centering
  \includegraphics[width=0.8\linewidth]{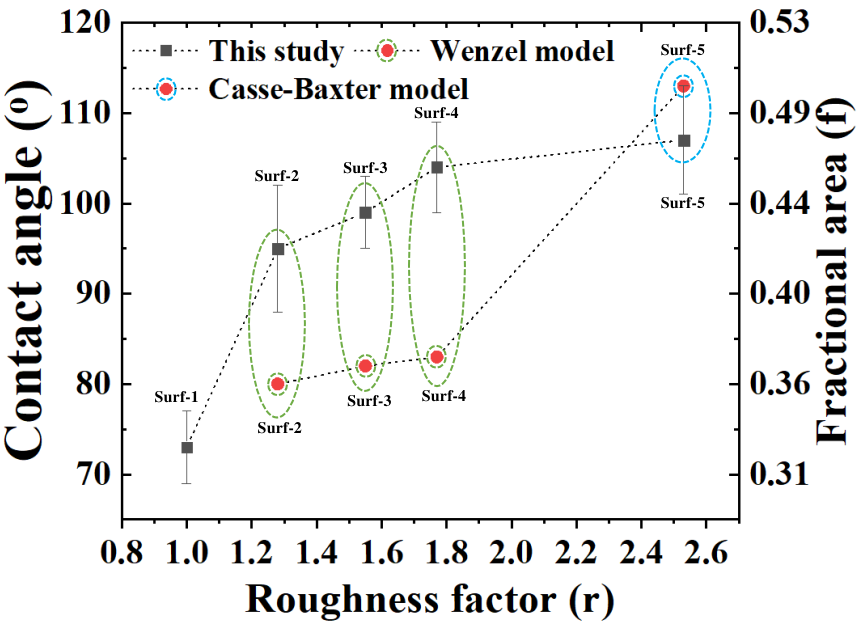}
  \caption{Comparison of Contact Angle (CA) predictions between MD simulations and classical wetting models (Wenzel and Cassie-Baxter) for Ti Surfaces with nanometric reliefs.}
  \label{fig:rfactor34}
\end{figure*}

Thus, in general,  the CA increases as roughness rises shifting the wettability of Ti surfaces from hydrophilic (Surf-1) to hydrophobic state for other surfaces. These observations for Surf-2, 3, and 4 align with the main assumption of the Wenzel model.  The Cassie-Baxter model somewhat better describes Surf-5, where air pockets exist between the droplet and the rough Ti surface, generally leading to higher CAs and less wetting. Interestingly, despite small sizes, the initial rapid decrease in CA and an increase in spreading radius is similar to the dynamic wetting theories and previous simulation studies, where the droplet adjusts its shape quickly upon initial contact to minimize the surface energy. Hence, the simulation model replicates key features of nanotextured surfaces created by femtosecond laser processing, using simplified nanostructures for better control and precision \citep{vorobyev2013direct, zhang2022femtosecond, gnilitskyi2017high}. While experimentally obtained laser-induced textures have some level of complexity and irregularity, the geometric parameters (e.g., height, width, spacing, and period) were based on experimental observations to ensure physical relevance . This approach allows us to focus on essential nanodroplet-surface interactions while maintaining computational efficiency and providing insights applicable to real-world conditions \citep{predictwetting}.

It is important to mention that the nanodroplet undergoes significant vibration on all the surfaces from 50 ps to 1000 ps, as seen in Fig. \ref{fig:Dy1}, \ref{fig:Sp} and \ref{fig:Dy}. The damping of droplet vibrations on rough surfaces is driven by several key mechanisms. As the surface roughness increases, the contact area between the droplet and surface also increases, leading to higher energy dissipation at the liquid-solid interface  \citep{sampayo2010communications, do2009static, gao2018nanodroplets}. Surfaces with deeper grooves (as observed in Surf-5) tend to trap air pockets beneath the droplet, which creates additional resistance against droplet movement. This increased resistance helps to dampen the vibrations more effectively compared to smoother surfaces. Conversely, surfaces like Surf-2 with moderate roughness facilitate higher spreading but lower vibration damping, as the droplet conforms more closely to the surface grooves without substantial energy dissipation from trapped air pockets.

The vibrations are also influenced by the droplet’s interaction with nanoscale surface features, where molecular-level forces, including van der Waals forces and electrostatic interactions, dominate \citep{song2013molecular}. The enhanced interaction between water molecules and the Ti surface atoms at nanometer roughness increases energy transfer from the droplet to the surface, contributing to the observed vibration damping \citep{malijevsky2012perspective}. The findings have significant implications for practical applications, particularly in the biomedical field where Ti surfaces are frequently used for implants \citep{raimbault2016effects}. The vibration damping effect observed with nanostructured surfaces improves the implant's performance by ensuring stable liquid interaction, thus improving biofluid adhesion and minimizing the risk of bacterial attachment, which is important for preventing infections.

\subsection{Static contact angle of nanodroplet on Ti surfaces}

It is important to highlight again that the nanodroplet on flat and rough Ti surfaces begins to reach a stable configuration at about 50 ps as shown in Fig. \ref{fig:Dy1}, \ref{fig:Sp}, and \ref{fig:Dy}. Therefore, this time scale is used to calculate the static CAs on the corresponding surfaces. The estimation of the static CAs of water nanodroplets on Ti surfaces enables the measurement of the surface wettability.

The determination of CA at the atomistic scale poses significant challenges since the nanodroplet boundaries undergo continuous changes influenced by molecular interactions, inherent fluctuations, droplet vibrations, shape changes, and surface roughness \citep{santiso2013calculation}. It is difficult to precisely define the nanodroplet shape at any specific moment. In the literature, the CA in MD studies is commonly estimated using a technique that relies on fitting the time-averaged density profile of the droplet \cite{chen2020mechanism, vsolc2011wettability, do2009static}.

 \begin{figure*}[t]
  \centering
  \includegraphics[width=1.02\linewidth]{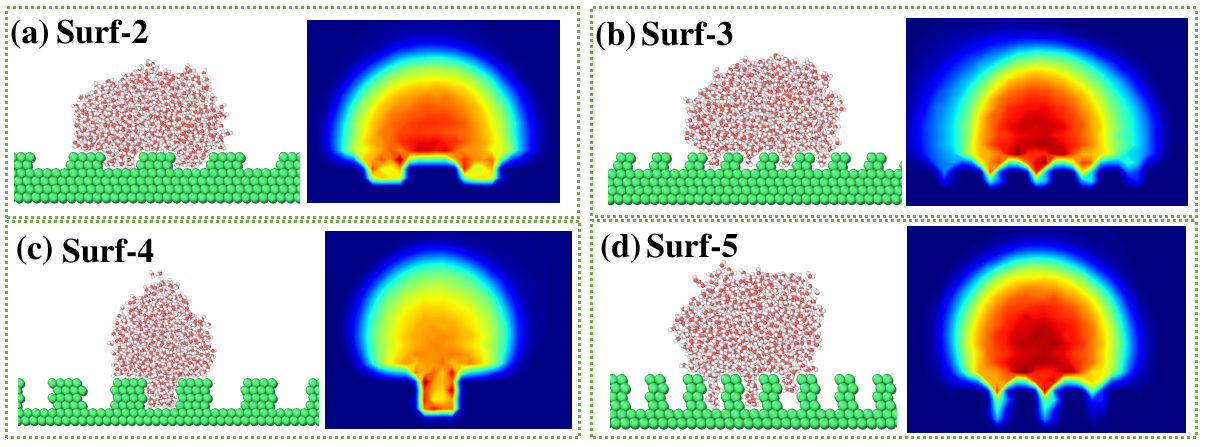}
  \caption{Snapshots of the equilibrium state and density profile of water nanodroplets on various Ti surfaces with different nanometric surface reliefs.}
  \label{fig:wetting_regim}
\end{figure*}

The approach adopted in measuring the static CA in this study involves using a circular fitting technique \citep{yaghoubi2018molecular} on the liquid-vapor interface of water nanodroplets on flat Ti surfaces, as illustrated by the density profile plots in Fig. \ref{fig:density} (a) and (b). Hence, Fig. \ref{fig:density} (a) shows the CA of a water nanodroplet on a flat Ti surface, and Fig. \ref{fig:density} (b) shows the density profile of the nanodroplet on the same surface. The density profile was generated by dividing the simulation domain into two-dimensional slices perpendicular to the y-z plane. To draw the density profile, the water density was computed for each slice with a size of 0.2 nm × 0.2 nm.

In the density profile, 1 corresponds to the liquid and 0 corresponds to the vapor. The liquid-vapor boundary was taken at the point where the density is half (0.5 $g/cm^3$) of the bulk of the water nanodroplet. Both visualizations are used to determine the static CA on flat Ti. The circular fitting was made in such a way that it passes above Ti surface such that near the wall density points of the droplet is omitted to avoid the influence from vibration at the liquid-solid interface \citep{barisik2013wetting}. The water CA was then determined using commercially available software \citep{schneider2012nih} to calculate the angle formed between the tangent line at the contact point on the liquid-vapor interface and the Ti surface. 

A series of $20$ snapshots were taken at time intervals of 50 ps. Taking this large number of snapshots at different points in the simulation allows us to average out the random fluctuations and obtain a more reliable estimate of the average CA. As such, the final static water CA was determined by averaging the measured values and accounting for the associated error. The snapshots of the CA calculation on flat Ti are summarized in the Supporting Information (Fig. 1.0). For the flat Ti surface, the estimated CA is $73^\circ \pm 4^\circ$, indicating that the flat Ti surface is strongly hydrophilic. Despite the size of the droplet being considerably smaller than those encountered in the experimental works \cite{cunha2016femtosecond, kim2021biological, liu2018unique, rad2014improved, goodarzi2016titanium}, the obtained CA value falls within the range of experimental observations on polished Ti. It is important to note that the CA in experiments can vary depending on the specific surface preparation and cleanliness of the Ti surface. Most of the CA values reported in the experimental literature for untreated Ti are $48.2^\circ$ \cite{cunha2016femtosecond}, $53.2^\circ$\cite{kim2021biological}, $56^\circ$ \cite{liu2018unique}, $66.4^\circ$ \citep{rad2014improved} and $75.5^\circ$ \citep{goodarzi2016titanium}. Consequently, the MD simulations faithfully reproduced the experimentally observed CAs for Ti surfaces, confirming the accuracy and reliability of the computational predictions of CA using an atomistic approach.

Several measurable macroscopic properties, such as CA, wettability, and surface roughness, are compared with the simulation results to establish consistency. It is, however, important to note that, direct comparison between MD simulations and experimental results faces certain challenges, particularly due to the differences in scale. The simulations in this study deal with nanodroplets, whereas most experimental studies use larger microscale droplets.

We note that the behavior of nanodroplets, like the one of nanoparticles, is more affected by their surface tension and can reveal remarkable features that are still far from being completely understood. Although the qualitative trends such as the transition from Wenzel to Cassie-Baxter states are similar across scales, quantitative discrepancies may arise due to size effects. Furthermore, experimental surfaces often have imperfections and irregularities that are not captured in idealized simulation models, potentially causing deviations in observed CA or droplet dynamics. Additionally, environmental factors such as temperature and pressure, which are tightly controlled in simulations, may vary in experimental conditions, introducing further variability.

In what follows, the static CA for Surf-2, 3, 4, and 5 were estimated using the same procedure. And the corresponding static CAs for these surfaces are $95^\circ \pm 7^\circ$, $99^\circ \pm 4^\circ$, $104^\circ \pm 5^\circ$ and $107^\circ \pm 6^\circ$ respectively. The introduced roughness amplifies the CA of the flat Ti surface from a hydrophilic to a hydrophobic state, consistent with earlier discussed droplet dynamics and trends generally observed in the literature.

\subsection{Effect of surface roughness on static contact angle}

Following the fact that the introduced roughness on Ti surfaces, Surf-2, 3, 4, and 5 changes the wetting conditions as earlier stated, classic wetting theories such as the Wenzel and Cassie-Baxter theories are used to predict the CA on the surfaces. The results are then compared with the MD numerical simulation to study the effect of the surface roughness on static CA on Ti surfaces. More details on the classical theories can be found in the "Supporting Information 1.2".

Fig. \ref{fig:rfactor34} shows the comparison of MD simulations in this study with those predicted by classical wetting models, specifically the Wenzel and Cassie-Baxter models, for Ti with nanometric surface reliefs. The flat Ti surface (Surf-1) serves as a baseline for understanding the wetting behavior without any influence of roughness. The CA observed in the MD simulations for this surface is close to the intrinsic Young CA. This value sets the stage for comparing how roughness modifies the wettability of the surface.

 The MD simulations show an increase in the static CA for Surf-2, 3, and 4 compared to the flat surface (Surf-1). This behavior shows that the droplet conforms to the surface roughness, increasing the actual contact area between the nanodroplet and the nanorough Ti surface. As a result of this, the wetting properties of the surfaces changed from a hydrophilic state to a hydrophobic state. The Wenzel model also predicted a higher CA compared to the flat surface, thus confirming the effect of the introduced relief. However, there is a disagreement between the MD calculations and the Wenzel model, as the measured CA in MD is greater than the CA predicted by the Wenzel model, with an observed deviation of about $15^\circ$, $17^\circ$ and $21^\circ$ respectively. Even though Surf-2, 3, and 4 do not entirely agree with the quantitative predictions of the Wenzel model, the qualitative assumptions of the model can still apply when the droplet fills the textured surfaces and no air pockets are present. 

As shown earlier for Surf-5, the water nanodroplet does not fully penetrate the surface roughness but instead rests on top of it, trapping air pockets underneath, thereby increasing the CA. Therefore, the CA observed in the MD simulations for Surf-5 is higher than that of the other four surfaces. The Cassie-Baxter model predicts an even higher CA than the MD results, indicating partial disagreement between the model and the MD simulations. Despite this disagreement, Surf-5 exhibits characteristics of the Cassie-Baxter regime, where the reduced solid-liquid contact area results in an increased CA and amplified hydrophobic effect. 

It is worthy of note that the observed deviations between the MD simulation CAs and the predictions of classical wetting models can be due to several factors that highlight the difficulties of wetting at the nanoscale. The classical models typically consider macroscopic droplets, while this study focuses on nanodroplets, where surface effects and molecular interactions become more pronounced. The increased surface-to-volume ratio of nanodroplets can lead to different wetting behavior compared to larger droplets, as the relative contributions of surface energy and line tension become more significant. Furthermore, the dynamic nature of wetting, with evolving droplet shapes and molecular rearrangements, may not be fully captured by static models like Wenzel and Cassie-Baxter. These models provide a snapshot of the wetting state at a particular time but do not account for the time-dependent evolution of the droplet shape and CA.

  \begin{figure}[h]
  \centering
  \includegraphics[width=1\linewidth]{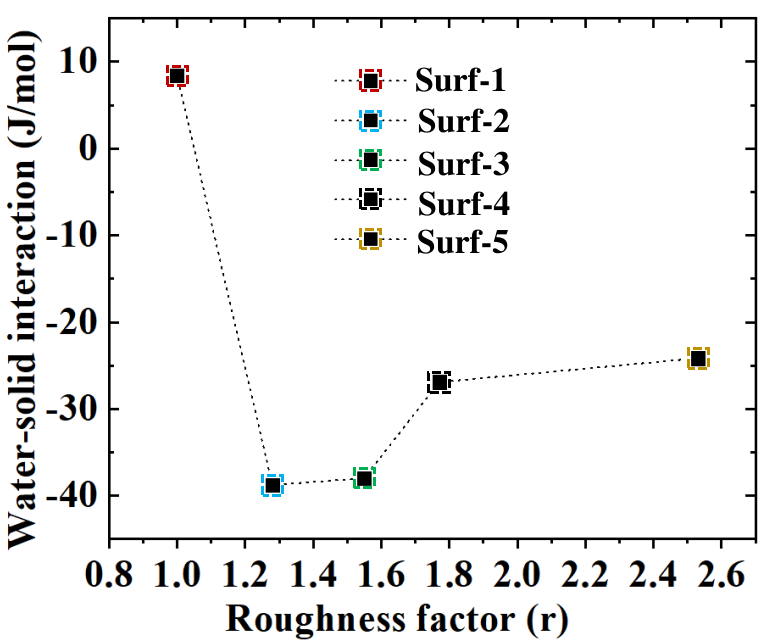}
  \caption{Water-solid interaction energy for flat and rough Ti surfaces with different roughness factors (r).}
  \label{fig:energy}
\end{figure}

To better confirm the wetting regime of the Ti surfaces described above, we compared the equilibrium state of the nanodroplet and the density profile plot, as shown in Fig. \ref{fig:wetting_regim}. This comparison provides a clear visualization of how the nanodroplet interacts with the various surface textures. Surf-2 and 4 exhibit a fully wetted state, indicating that the droplets adapt to the shape and contours of the surface roughness. In addition, Surf-2, for instance, fills two of the surface grooves, resulting in a lower CA compared to Surf-4. Surf-4, although filling only one groove, shows a higher CA than Surf-2. This behavior supports the Wenzel regime, where the droplet’s CA is influenced by the increased surface area due to the roughness.

In the case of Surf-3 and 5, the wetting trends are similar but distinct from Surf-2 and 4. On Surf-3, the water droplet fully occupies the grooves of the surface textures, consistent with what is observed in the Wenzel wetting regime. Surf-5, however, demonstrates a different interaction. The nanodroplet in Surf-5 partially fills the grooves, similar to Surf-3, but leaves portions of the relief empty, creating air pockets beneath the droplet. This partial filling and presence of air pockets lead to a higher CA, consistent with the Cassie-Baxter regime. 

Therefore, Surf-5 represents the Cassie-Baxter state, where the surface roughness promotes the formation of a bi-phase air-liquid layer at the interface thus increasing hydrophobicity. On the other hand, Surf-2, 3, and 4 exhibit characteristics of the Wenzel state, where the droplet fully wets the rough surface, leading to different degrees of wettability as indicated by their CAs. This analysis further confirms the claim in the dynamic and spreading process of water nanodroplet on different Ti surfaces.

Fig. \ref{fig:energy} illustrates the relationship between the water-solid interaction energy and the roughness factor (r) of Ti surfaces. This interaction energy, defined as the total Van der Waals interaction between water molecules and Ti surfaces is important in understanding the behavior of nanodroplet in contact with varying surface roughness on Ti. Specifically, we evaluated the water-solid interaction energy for each surface to quantify how roughness influences wetting behavior. In addition to several analytical models that assumes that only shape and the slope of the pillar edge are responsible for the unusual water droplet spreading, we showed that surface energy effects are in line with previous experimental and theoretical observations \citep{alen2016analytical, hay2008theoretical, zhang2016molecular}. In particular, these calculations help to clarify the correlation between surface roughness and the observed transitions between the Wenzel and Cassie-Baxter wetting regimes. This approach will provide a more comprehensive understanding of how nanostructured surfaces affect energy dissipation during the wetting process.

For the flat surface (Surf-1), the interaction energy is positive, which indicates a net repulsive force between water molecules and the flat Ti surface leading to weaker adhesion properties with nanodroplet due to smaller surface area for interaction. In Surf-2, the interaction energy is negative. This sharp drop reveals strong adhesive interactions, which lead to wetting of the rough Ti surface by the nanodroplet. This phenomenon is due to the increased surface area and roughness that increased the Van der Waals interactions between the water molecules and the rough Ti surface. This further gives more validation to why the spreading radius of Surf-2 is significantly higher than the other surfaces as previously discussed. Surf-3 shows a slight reduction in the magnitude of interaction energy. This demonstrates that the highly attractive interactions noted in Surf-2 are slightly reduced as roughness continues to increase. 

In Surf-4, the interaction energy becomes less negative, indicating weaker interactions. This trend continues with Surf-5, where the interaction energy is even less negative compared to Surf-4. This demonstrates weaker interactions, leading to non-wetting behavior. These observations show that as the roughness increases, the attractive interactions between the nanodroplet and the Ti surface diminish, making the surface more hydrophobic. This occurs because the nanodroplet does not fully conform to the rough surface, resulting in fewer contact points for Van der Waals interactions. Further increasing the roughness reduces the magnitude of these attractive forces. This finding is consistent with trends in the literature, which show that when the droplet state transitions from the Wenzel state to the Cassie state, the interaction energy decreases sharply \citep{niu2014static, zhao2021molecular, zhang2016molecular}. Therefore, understanding and controlling surface roughness can play an important role in tailoring the interfacial properties of Ti surfaces for specific applications. 

\section{Conclusions}

This study provides atomic-level insights into the wetting mechanisms of water nanodroplets on both flat and nano-textured titanium surfaces, using molecular dynamics simulations, with implications for medical implants. Our results demonstrate a clear correlation between surface roughness and wettability, with higher roughness promoting a transition from Wenzel to Cassie-Baxter wetting regimes. The simulation results also show that the introduction of surface roughness significantly influences the spreading behavior of the nanodroplet, with larger grooves promoting extensive spreading and narrower grooves limiting it. Furthermore, we observed the damping of droplet vibrations and a transition to a hydrophobic state on both flat and rough surfaces, driven by atomic bonding mechanisms specific to Ti surfaces. The MD simulations reveal interaction energy trends that show strong adhesive interactions on rough surfaces, leading to increased hydrophobicity as roughness further increases. The discrepancies between simulation contact angle and classical wetting models shows the importance of considering nanoscale effects and molecular-level interactions in understanding and predicting wetting phenomena. These findings have significant implications for the design and optimization of titanium-based biomaterials, as well as for the development of advanced materials with tailored wetting properties for various applications, including self-cleaning surfaces and microfluidic devices.


\section*{Supporting Information}
The Supporting Information is available free of charge at [Link].

Calculated structural properties of titanium, snapshots of water nanodroplets on flat titanium surface, and classical wetting models, including the Wenzel and Cassie-Baxter models [PDF].

\section*{Declaration of competing interest}
The authors declare no conflicts of interest.

\section*{Acknowledgments}
The research was funded by the French Ministère de l'Éducation Nationale and the FET Laser Implant project, which is part of the EU HORIZON 2020 program, under Grant Agreement ID: 951730.  We wish to acknowledge the support provided in part by EUR Manutech Sleight and Labex Manutech-SISE. We extend our gratitude for the computational support provided by Hubert Curien Laboratory and CINES, France, referenced under project number AD010814604R1. 



\bibliographystyle{pisikabst}
\bibliography{bibfile}

\begin{thebibliography}{67}
\expandafter\ifx\csname natexlab\endcsname\relax\def\natexlab#1{#1}\fi
\expandafter\ifx\csname bibnamefont\endcsname\relax
  \def\bibnamefont#1{#1}\fi
\expandafter\ifx\csname bibfnamefont\endcsname\relax
  \def\bibfnamefont#1{#1}\fi
\expandafter\ifx\csname citenamefont\endcsname\relax
  \def\citenamefont#1{#1}\fi
\expandafter\ifx\csname url\endcsname\relax
  \def\url#1{\texttt{#1}}\fi
\expandafter\ifx\csname urlprefix\endcsname\relax\def\urlprefix{}\fi
\providecommand{\bibinfo}[2]{#2}
\providecommand{\eprint}[1]{\href{http://arxiv.org/abs/#1}{arXiv:#1}}

\bibitem[{\citenamefont{Elias et~al.}(2008)\citenamefont{Elias, Lima, Valiev, and Meyers}}]{elias2008biomedical}
\bibinfo{author}{\bibfnamefont{Elias} \bibnamefont{C.}}, \bibinfo{author}{\bibfnamefont{Lima} \bibnamefont{J.}}, \bibinfo{author}{\bibfnamefont{Valiev} \bibnamefont{R.}}, \bibnamefont{and} \bibinfo{author}{\bibfnamefont{Meyers} \bibnamefont{M.}}, \bibinfo{title}{Biomedical applications of titanium and its alloys}, \emph{\bibinfo{journal}{Jom}} \textbf{\bibinfo{volume}{60}}, \bibinfo{pages}{46} (\bibinfo{year}{2008}).

\bibitem[{\citenamefont{Kligman et~al.}(2021)\citenamefont{Kligman, Ren, Chung, Perillo, Chang, Koo, Zheng, and Li}}]{kligman2021impact}
\bibinfo{author}{\bibfnamefont{Kligman} \bibnamefont{S.}}, \bibinfo{author}{\bibfnamefont{Ren} \bibnamefont{Z.}}, \bibinfo{author}{\bibfnamefont{Chung} \bibnamefont{C.-H.}}, \bibinfo{author}{\bibfnamefont{Perillo} \bibnamefont{M.~A.}}, \bibinfo{author}{\bibfnamefont{Chang} \bibnamefont{Y.-C.}}, \bibinfo{author}{\bibfnamefont{Koo} \bibnamefont{H.}}, \bibinfo{author}{\bibfnamefont{Zheng} \bibnamefont{Z.}}, \bibnamefont{and} \bibinfo{author}{\bibfnamefont{Li}~\bibnamefont{C.}}, \bibinfo{title}{The impact of dental implant surface modifications on osseointegration and biofilm formation}, \emph{\bibinfo{journal}{Journal of clinical medicine}} \textbf{\bibinfo{volume}{10}}, \bibinfo{pages}{1641} (\bibinfo{year}{2021}).

\bibitem[{\citenamefont{Lutey et~al.}(2018)\citenamefont{Lutey, Gemini, Romoli, Lazzini, Fuso, Faucon, and Kling}}]{lutey2018towards}
\bibinfo{author}{\bibfnamefont{Lutey} \bibnamefont{A.~H.}}, \bibinfo{author}{\bibfnamefont{Gemini} \bibnamefont{L.}}, \bibinfo{author}{\bibfnamefont{Romoli} \bibnamefont{L.}}, \bibinfo{author}{\bibfnamefont{Lazzini} \bibnamefont{G.}}, \bibinfo{author}{\bibfnamefont{Fuso} \bibnamefont{F.}}, \bibinfo{author}{\bibfnamefont{Faucon} \bibnamefont{M.}}, \bibnamefont{and} \bibinfo{author}{\bibfnamefont{Kling} \bibnamefont{R.}}, \bibinfo{title}{Towards laser-textured antibacterial surfaces}, \emph{\bibinfo{journal}{Scientific reports}} \textbf{\bibinfo{volume}{8}}, \bibinfo{pages}{10112} (\bibinfo{year}{2018}).

\bibitem[{\citenamefont{Vorobyev and Guo}(2013)}]{vorobyev2013direct}
\bibinfo{author}{\bibfnamefont{Vorobyev} \bibnamefont{A.~Y.}} \bibnamefont{and} \bibinfo{author}{\bibfnamefont{Guo} \bibnamefont{C.}}, \bibinfo{title}{Direct femtosecond laser surface nano/microstructuring and its applications}, \emph{\bibinfo{journal}{Laser \& Photonics Reviews}} \textbf{\bibinfo{volume}{7}}, \bibinfo{pages}{385} (\bibinfo{year}{2013}).

\bibitem[{\citenamefont{Yong et~al.}(2018)\citenamefont{Yong, Chen, Yang, Jiang, and Hou}}]{yong2018hall}
\bibinfo{author}{\bibfnamefont{Yong} \bibnamefont{J.}}, \bibinfo{author}{\bibfnamefont{Chen} \bibnamefont{F.}}, \bibinfo{author}{\bibfnamefont{Yang} \bibnamefont{Q.}}, \bibinfo{author}{\bibfnamefont{Jiang} \bibnamefont{Z.}}, \bibnamefont{and} \bibinfo{author}{\bibfnamefont{Hou} \bibnamefont{X.}}, \bibinfo{title}{Hall of fame article: A review of femtosecond-laser-induced underwater superoleophobic surfaces (adv. mater. interfaces 7/2018)}, \emph{\bibinfo{journal}{Advanced Materials Interfaces}} \textbf{\bibinfo{volume}{5}}, \bibinfo{pages}{1870033} (\bibinfo{year}{2018}).

\bibitem[{\citenamefont{Sugioka and Cheng}(2014)}]{sugioka2014ultrafast}
\bibinfo{author}{\bibfnamefont{Sugioka} \bibnamefont{K.}} \bibnamefont{and} \bibinfo{author}{\bibfnamefont{Cheng} \bibnamefont{Y.}}, \bibinfo{title}{Ultrafast lasers—reliable tools for advanced materials processing}, \emph{\bibinfo{journal}{Light: Science \& Applications}} \textbf{\bibinfo{volume}{3}}, \bibinfo{pages}{e149} (\bibinfo{year}{2014}).

\bibitem[{\citenamefont{Gittens et~al.}(2014)\citenamefont{Gittens, Scheideler, Rupp, Hyzy, Geis-Gerstorfer, Schwartz, and Boyan}}]{gittens2014review}
\bibinfo{author}{\bibfnamefont{Gittens} \bibnamefont{R.~A.}}, \bibinfo{author}{\bibfnamefont{Scheideler} \bibnamefont{L.}}, \bibinfo{author}{\bibfnamefont{Rupp} \bibnamefont{F.}}, \bibinfo{author}{\bibfnamefont{Hyzy} \bibnamefont{S.~L.}}, \bibinfo{author}{\bibfnamefont{Geis-Gerstorfer} \bibnamefont{J.}}, \bibinfo{author}{\bibfnamefont{Schwartz} \bibnamefont{Z.}}, \bibnamefont{and} \bibinfo{author}{\bibfnamefont{Boyan} \bibnamefont{B.~D.}}, \bibinfo{title}{A review on the wettability of dental implant surfaces ii: Biological and clinical aspects}, \emph{\bibinfo{journal}{Acta biomaterialia}} \textbf{\bibinfo{volume}{10}}, \bibinfo{pages}{2907} (\bibinfo{year}{2014}).

\bibitem[{\citenamefont{Qu{\'e}r{\'e}}(2008)}]{quere2008wetting}
\bibinfo{author}{\bibfnamefont{Qu{\'e}r{\'e}} \bibnamefont{D.}}, \bibinfo{title}{Wetting and roughness}, \emph{\bibinfo{journal}{Annu. Rev. Mater. Res.}} \textbf{\bibinfo{volume}{38}}, \bibinfo{pages}{71} (\bibinfo{year}{2008}).

\bibitem[{\citenamefont{Wu et~al.}(2009)\citenamefont{Wu, Zhou, Li, Ye, Li, and Cai}}]{wu2009superhydrophobic}
\bibinfo{author}{\bibfnamefont{Wu}~\bibnamefont{B.}}, \bibinfo{author}{\bibfnamefont{Zhou} \bibnamefont{M.}}, \bibinfo{author}{\bibfnamefont{Li}~\bibnamefont{J.}}, \bibinfo{author}{\bibfnamefont{Ye}~\bibnamefont{X.}}, \bibinfo{author}{\bibfnamefont{Li}~\bibnamefont{G.}}, \bibnamefont{and} \bibinfo{author}{\bibfnamefont{Cai} \bibnamefont{L.}}, \bibinfo{title}{Superhydrophobic surfaces fabricated by microstructuring of stainless steel using a femtosecond laser}, \emph{\bibinfo{journal}{Applied surface science}} \textbf{\bibinfo{volume}{256}}, \bibinfo{pages}{61} (\bibinfo{year}{2009}).

\bibitem[{\citenamefont{Cunha et~al.}(2013)\citenamefont{Cunha, Serro, Oliveira, Almeida, Vilar, and Durrieu}}]{cunha2013wetting}
\bibinfo{author}{\bibfnamefont{Cunha} \bibnamefont{A.}}, \bibinfo{author}{\bibfnamefont{Serro} \bibnamefont{A.~P.}}, \bibinfo{author}{\bibfnamefont{Oliveira} \bibnamefont{V.}}, \bibinfo{author}{\bibfnamefont{Almeida} \bibnamefont{A.}}, \bibinfo{author}{\bibfnamefont{Vilar} \bibnamefont{R.}}, \bibnamefont{and} \bibinfo{author}{\bibfnamefont{Durrieu} \bibnamefont{M.-C.}}, \bibinfo{title}{Wetting behaviour of femtosecond laser textured ti--6al--4v surfaces}, \emph{\bibinfo{journal}{Applied Surface Science}} \textbf{\bibinfo{volume}{265}}, \bibinfo{pages}{688} (\bibinfo{year}{2013}).

\bibitem[{\citenamefont{Raimbault et~al.}(2016)\citenamefont{Raimbault, Benayoun, Anselme, Mauclair, Bourgade, Kietzig, Girard-Lauriault, Valette, and Donnet}}]{raimbault2016effects}
\bibinfo{author}{\bibfnamefont{Raimbault} \bibnamefont{O.}}, \bibinfo{author}{\bibfnamefont{Benayoun} \bibnamefont{S.}}, \bibinfo{author}{\bibfnamefont{Anselme} \bibnamefont{K.}}, \bibinfo{author}{\bibfnamefont{Mauclair} \bibnamefont{C.}}, \bibinfo{author}{\bibfnamefont{Bourgade} \bibnamefont{T.}}, \bibinfo{author}{\bibfnamefont{Kietzig} \bibnamefont{A.-M.}}, \bibinfo{author}{\bibfnamefont{Girard-Lauriault} \bibnamefont{P.-L.}}, \bibinfo{author}{\bibfnamefont{Valette} \bibnamefont{S.}}, \bibnamefont{and} \bibinfo{author}{\bibfnamefont{Donnet} \bibnamefont{C.}}, \bibinfo{title}{The effects of femtosecond laser-textured ti-6al-4v on wettability and cell response}, \emph{\bibinfo{journal}{Materials Science and Engineering: C}} \textbf{\bibinfo{volume}{69}}, \bibinfo{pages}{311} (\bibinfo{year}{2016}).

\bibitem[{\citenamefont{Dou et~al.}(2020)\citenamefont{Dou, Liu, Xu, Chen, Miao, L{\"u}, and Jiang}}]{dou2020influence}
\bibinfo{author}{\bibfnamefont{Dou} \bibnamefont{H.-q.}}, \bibinfo{author}{\bibfnamefont{Liu} \bibnamefont{H.}}, \bibinfo{author}{\bibfnamefont{Xu}~\bibnamefont{S.}}, \bibinfo{author}{\bibfnamefont{Chen} \bibnamefont{Y.}}, \bibinfo{author}{\bibfnamefont{Miao} \bibnamefont{X.}}, \bibinfo{author}{\bibfnamefont{L{\"u}}~\bibnamefont{H.}}, \bibnamefont{and} \bibinfo{author}{\bibfnamefont{Jiang} \bibnamefont{X.}}, \bibinfo{title}{Influence of laser fluences and scan speeds on the morphologies and wetting properties of titanium alloy}, \emph{\bibinfo{journal}{Optik}} \textbf{\bibinfo{volume}{224}}, \bibinfo{pages}{165443} (\bibinfo{year}{2020}).

\bibitem[{\citenamefont{Wenzel}(1936)}]{wenzel1936resistance}
\bibinfo{author}{\bibfnamefont{Wenzel} \bibnamefont{R.~N.}}, \bibinfo{title}{Resistance of solid surfaces to wetting by water}, \emph{\bibinfo{journal}{Industrial \& engineering chemistry}} \textbf{\bibinfo{volume}{28}}, \bibinfo{pages}{988} (\bibinfo{year}{1936}).

\bibitem[{\citenamefont{Cassie and Baxter}(1944)}]{cassie1944wettability}
\bibinfo{author}{\bibfnamefont{Cassie} \bibnamefont{A.}} \bibnamefont{and} \bibinfo{author}{\bibfnamefont{Baxter} \bibnamefont{S.}}, \bibinfo{title}{Wettability of porous surfaces}, \emph{\bibinfo{journal}{Transactions of the Faraday society}} \textbf{\bibinfo{volume}{40}}, \bibinfo{pages}{546} (\bibinfo{year}{1944}).

\bibitem[{\citenamefont{Bizi-Bandoki et~al.}(2011)\citenamefont{Bizi-Bandoki, Benayoun, Valette, Beaugiraud, and Audouard}}]{bizi2011modifications}
\bibinfo{author}{\bibfnamefont{Bizi-Bandoki} \bibnamefont{P.}}, \bibinfo{author}{\bibfnamefont{Benayoun} \bibnamefont{S.}}, \bibinfo{author}{\bibfnamefont{Valette} \bibnamefont{S.}}, \bibinfo{author}{\bibfnamefont{Beaugiraud} \bibnamefont{B.}}, \bibnamefont{and} \bibinfo{author}{\bibfnamefont{Audouard} \bibnamefont{E.}}, \bibinfo{title}{Modifications of roughness and wettability properties of metals induced by femtosecond laser treatment}, \emph{\bibinfo{journal}{Applied Surface Science}} \textbf{\bibinfo{volume}{257}}, \bibinfo{pages}{5213} (\bibinfo{year}{2011}).

\bibitem[{\citenamefont{Yong et~al.}(2014)\citenamefont{Yong, Yang, Chen, Zhang, Farooq, Du, and Hou}}]{yong2014simple}
\bibinfo{author}{\bibfnamefont{Yong} \bibnamefont{J.}}, \bibinfo{author}{\bibfnamefont{Yang} \bibnamefont{Q.}}, \bibinfo{author}{\bibfnamefont{Chen} \bibnamefont{F.}}, \bibinfo{author}{\bibfnamefont{Zhang} \bibnamefont{D.}}, \bibinfo{author}{\bibfnamefont{Farooq} \bibnamefont{U.}}, \bibinfo{author}{\bibfnamefont{Du}~\bibnamefont{G.}}, \bibnamefont{and} \bibinfo{author}{\bibfnamefont{Hou} \bibnamefont{X.}}, \bibinfo{title}{A simple way to achieve superhydrophobicity, controllable water adhesion, anisotropic sliding, and anisotropic wetting based on femtosecond-laser-induced line-patterned surfaces}, \emph{\bibinfo{journal}{Journal of Materials Chemistry A}} \textbf{\bibinfo{volume}{2}}, \bibinfo{pages}{5499} (\bibinfo{year}{2014}).

\bibitem[{\citenamefont{Giannuzzi et~al.}(2019)\citenamefont{Giannuzzi, Gaudiuso, Di~Mundo, Mirenghi, Fraggelakis, Kling, Lugar{\`a}, and Ancona}}]{giannuzzi2019short}
\bibinfo{author}{\bibfnamefont{Giannuzzi} \bibnamefont{G.}}, \bibinfo{author}{\bibfnamefont{Gaudiuso} \bibnamefont{C.}}, \bibinfo{author}{\bibfnamefont{Di~Mundo} \bibnamefont{R.}}, \bibinfo{author}{\bibfnamefont{Mirenghi} \bibnamefont{L.}}, \bibinfo{author}{\bibfnamefont{Fraggelakis} \bibnamefont{F.}}, \bibinfo{author}{\bibfnamefont{Kling} \bibnamefont{R.}}, \bibinfo{author}{\bibfnamefont{Lugar{\`a}} \bibnamefont{P.~M.}}, \bibnamefont{and} \bibinfo{author}{\bibfnamefont{Ancona} \bibnamefont{A.}}, \bibinfo{title}{Short and long term surface chemistry and wetting behaviour of stainless steel with 1d and 2d periodic structures induced by bursts of femtosecond laser pulses}, \emph{\bibinfo{journal}{Applied Surface Science}} \textbf{\bibinfo{volume}{494}}, \bibinfo{pages}{1055} (\bibinfo{year}{2019}).

\bibitem[{\citenamefont{Cunha et~al.}(2016)\citenamefont{Cunha, Elie, Plawinski, Serro, do~Rego, Almeida, Urdaci, Durrieu, and Vilar}}]{cunha2016femtosecond}
\bibinfo{author}{\bibfnamefont{Cunha} \bibnamefont{A.}}, \bibinfo{author}{\bibfnamefont{Elie} \bibnamefont{A.-M.}}, \bibinfo{author}{\bibfnamefont{Plawinski} \bibnamefont{L.}}, \bibinfo{author}{\bibfnamefont{Serro} \bibnamefont{A.~P.}}, \bibinfo{author}{\bibfnamefont{do~Rego} \bibnamefont{A.~M.~B.}}, \bibinfo{author}{\bibfnamefont{Almeida} \bibnamefont{A.}}, \bibinfo{author}{\bibfnamefont{Urdaci} \bibnamefont{M.~C.}}, \bibinfo{author}{\bibfnamefont{Durrieu} \bibnamefont{M.-C.}}, \bibnamefont{and} \bibinfo{author}{\bibfnamefont{Vilar} \bibnamefont{R.}}, \bibinfo{title}{Femtosecond laser surface texturing of titanium as a method to reduce the adhesion of staphylococcus aureus and biofilm formation}, \emph{\bibinfo{journal}{Applied Surface Science}} \textbf{\bibinfo{volume}{360}}, \bibinfo{pages}{485} (\bibinfo{year}{2016}).

\bibitem[{\citenamefont{Dumas et~al.}(2012)\citenamefont{Dumas, Rattner, Vico, Audouard, Dumas, Naisson, and Bertrand}}]{dumas2012multiscale}
\bibinfo{author}{\bibfnamefont{Dumas} \bibnamefont{V.}}, \bibinfo{author}{\bibfnamefont{Rattner} \bibnamefont{A.}}, \bibinfo{author}{\bibfnamefont{Vico} \bibnamefont{L.}}, \bibinfo{author}{\bibfnamefont{Audouard} \bibnamefont{E.}}, \bibinfo{author}{\bibfnamefont{Dumas} \bibnamefont{J.~C.}}, \bibinfo{author}{\bibfnamefont{Naisson} \bibnamefont{P.}}, \bibnamefont{and} \bibinfo{author}{\bibfnamefont{Bertrand} \bibnamefont{P.}}, \bibinfo{title}{Multiscale grooved titanium processed with femtosecond laser influences mesenchymal stem cell morphology, adhesion, and matrix organization}, \emph{\bibinfo{journal}{Journal of Biomedical Materials Research Part A}} \textbf{\bibinfo{volume}{100}}, \bibinfo{pages}{3108} (\bibinfo{year}{2012}).

\bibitem[{\citenamefont{Sirdeshmukh and Dongre}(2021)}]{sirdeshmukh2021laser}
\bibinfo{author}{\bibfnamefont{Sirdeshmukh} \bibnamefont{N.}} \bibnamefont{and} \bibinfo{author}{\bibfnamefont{Dongre} \bibnamefont{G.}}, \bibinfo{title}{Laser micro \& nano surface texturing for enhancing osseointegration and antimicrobial effect of biomaterials: A review}, \emph{\bibinfo{journal}{Materials Today: Proceedings}} \textbf{\bibinfo{volume}{44}}, \bibinfo{pages}{2348} (\bibinfo{year}{2021}).

\bibitem[{\citenamefont{Shivakoti et~al.}(2021)\citenamefont{Shivakoti, Kibria, Cep, Pradhan, and Sharma}}]{shivakoti2021laser}
\bibinfo{author}{\bibfnamefont{Shivakoti} \bibnamefont{I.}}, \bibinfo{author}{\bibfnamefont{Kibria} \bibnamefont{G.}}, \bibinfo{author}{\bibfnamefont{Cep} \bibnamefont{R.}}, \bibinfo{author}{\bibfnamefont{Pradhan} \bibnamefont{B.~B.}}, \bibnamefont{and} \bibinfo{author}{\bibfnamefont{Sharma} \bibnamefont{A.}}, \bibinfo{title}{Laser surface texturing for biomedical applications: a review}, \emph{\bibinfo{journal}{Coatings}} \textbf{\bibinfo{volume}{11}}, \bibinfo{pages}{124} (\bibinfo{year}{2021}).

\bibitem[{\citenamefont{Mukherjee et~al.}(2015)\citenamefont{Mukherjee, Dhara, and Saha}}]{mukherjee2015enhancing}
\bibinfo{author}{\bibfnamefont{Mukherjee} \bibnamefont{S.}}, \bibinfo{author}{\bibfnamefont{Dhara} \bibnamefont{S.}}, \bibnamefont{and} \bibinfo{author}{\bibfnamefont{Saha} \bibnamefont{P.}}, \bibinfo{title}{Enhancing the biocompatibility of ti6al4v implants by laser surface microtexturing: an in vitro study}, \emph{\bibinfo{journal}{The International Journal of Advanced Manufacturing Technology}} \textbf{\bibinfo{volume}{76}}, \bibinfo{pages}{5} (\bibinfo{year}{2015}).

\bibitem[{\citenamefont{Park and Aluru}(2009)}]{park2009temperature}
\bibinfo{author}{\bibfnamefont{Park} \bibnamefont{J.~H.}} \bibnamefont{and} \bibinfo{author}{\bibfnamefont{Aluru} \bibnamefont{N.}}, \bibinfo{title}{Temperature-dependent wettability on a titanium dioxide surface}, \emph{\bibinfo{journal}{Molecular Simulation}} \textbf{\bibinfo{volume}{35}}, \bibinfo{pages}{31} (\bibinfo{year}{2009}).

\bibitem[{\citenamefont{Ohler and Langel}(2009)}]{ohler2009molecular}
\bibinfo{author}{\bibfnamefont{Ohler} \bibnamefont{B.}} \bibnamefont{and} \bibinfo{author}{\bibfnamefont{Langel} \bibnamefont{W.}}, \bibinfo{title}{Molecular dynamics simulations on the interface between titanium dioxide and water droplets: A new model for the contact angle}, \emph{\bibinfo{journal}{The Journal of Physical Chemistry C}} \textbf{\bibinfo{volume}{113}}, \bibinfo{pages}{10189} (\bibinfo{year}{2009}).

\bibitem[{\citenamefont{Zhu et~al.}(2023)\citenamefont{Zhu, Dastan, Liu, Wu, Shi, Chu, Altaf, and Mohammed}}]{zhu2023surface}
\bibinfo{author}{\bibfnamefont{Zhu} \bibnamefont{P.}}, \bibinfo{author}{\bibfnamefont{Dastan} \bibnamefont{D.}}, \bibinfo{author}{\bibfnamefont{Liu} \bibnamefont{L.}}, \bibinfo{author}{\bibfnamefont{Wu}~\bibnamefont{L.}}, \bibinfo{author}{\bibfnamefont{Shi} \bibnamefont{Z.}}, \bibinfo{author}{\bibfnamefont{Chu} \bibnamefont{Q.-Q.}}, \bibinfo{author}{\bibfnamefont{Altaf} \bibnamefont{F.}}, \bibnamefont{and} \bibinfo{author}{\bibfnamefont{Mohammed} \bibnamefont{M.~K.}}, \bibinfo{title}{Surface wettability of various phases of titania thin films: Atomic-scale simulation studies}, \emph{\bibinfo{journal}{Journal of Molecular Graphics and Modelling}} \textbf{\bibinfo{volume}{118}}, \bibinfo{pages}{108335} (\bibinfo{year}{2023}).

\bibitem[{\citenamefont{Berendsen et~al.}(1987)\citenamefont{Berendsen, Grigera, and Straatsma}}]{berendsen1987missing}
\bibinfo{author}{\bibfnamefont{Berendsen} \bibnamefont{H.~J.}}, \bibinfo{author}{\bibfnamefont{Grigera} \bibnamefont{J.~R.}}, \bibnamefont{and} \bibinfo{author}{\bibfnamefont{Straatsma} \bibnamefont{T.~P.}}, \bibinfo{title}{The missing term in effective pair potentials}, \emph{\bibinfo{journal}{Journal of Physical Chemistry}} \textbf{\bibinfo{volume}{91}}, \bibinfo{pages}{6269} (\bibinfo{year}{1987}).

\bibitem[{\citenamefont{Chen et~al.}(2020)\citenamefont{Chen, Wang, Xiang, and Tao}}]{chen2020mechanism}
\bibinfo{author}{\bibfnamefont{Chen} \bibnamefont{L.}}, \bibinfo{author}{\bibfnamefont{Wang} \bibnamefont{S.-Y.}}, \bibinfo{author}{\bibfnamefont{Xiang} \bibnamefont{X.}}, \bibnamefont{and} \bibinfo{author}{\bibfnamefont{Tao} \bibnamefont{W.-Q.}}, \bibinfo{title}{Mechanism of surface nanostructure changing wettability: A molecular dynamics simulation}, \emph{\bibinfo{journal}{Computational Materials Science}} \textbf{\bibinfo{volume}{171}}, \bibinfo{pages}{109223} (\bibinfo{year}{2020}).

\bibitem[{\citenamefont{Ryckaert et~al.}(1977)\citenamefont{Ryckaert, Ciccotti, and Berendsen}}]{ryckaert1977numerical}
\bibinfo{author}{\bibfnamefont{Ryckaert} \bibnamefont{J.-P.}}, \bibinfo{author}{\bibfnamefont{Ciccotti} \bibnamefont{G.}}, \bibnamefont{and} \bibinfo{author}{\bibfnamefont{Berendsen} \bibnamefont{H.~J.}}, \bibinfo{title}{Numerical integration of the cartesian equations of motion of a system with constraints: molecular dynamics of n-alkanes}, \emph{\bibinfo{journal}{Journal of computational physics}} \textbf{\bibinfo{volume}{23}}, \bibinfo{pages}{327} (\bibinfo{year}{1977}).

\bibitem[{\citenamefont{Hockney and Eastwood}(2021)}]{hockney2021computer}
\bibinfo{author}{\bibfnamefont{Hockney} \bibnamefont{R.~W.}} \bibnamefont{and} \bibinfo{author}{\bibfnamefont{Eastwood} \bibnamefont{J.~W.}}, \emph{\bibinfo{title}{Computer simulation using particles}} (\bibinfo{publisher}{crc Press}, \bibinfo{year}{2021}).

\bibitem[{\citenamefont{Darden et~al.}(1993)\citenamefont{Darden, York, and Pedersen}}]{darden1993particle}
\bibinfo{author}{\bibfnamefont{Darden} \bibnamefont{T.}}, \bibinfo{author}{\bibfnamefont{York} \bibnamefont{D.}}, \bibnamefont{and} \bibinfo{author}{\bibfnamefont{Pedersen} \bibnamefont{L.}}, \bibinfo{title}{Particle mesh ewald: An n\(\cdot\)log(n) method for ewald sums in large systems}, \emph{\bibinfo{journal}{The Journal of Chemical Physics}} \textbf{\bibinfo{volume}{98}}, \bibinfo{pages}{10089} (\bibinfo{year}{1993}).

\bibitem[{\citenamefont{Barisik and Beskok}(2013)}]{barisik2013wetting}
\bibinfo{author}{\bibfnamefont{Barisik} \bibnamefont{M.}} \bibnamefont{and} \bibinfo{author}{\bibfnamefont{Beskok} \bibnamefont{A.}}, \bibinfo{title}{Wetting characterisation of silicon (1, 0, 0) surface}, \emph{\bibinfo{journal}{Molecular Simulation}} \textbf{\bibinfo{volume}{39}}, \bibinfo{pages}{700} (\bibinfo{year}{2013}).

\bibitem[{\citenamefont{Allen et~al.}(1987)\citenamefont{Allen, Tildesley et~al.}}]{allen1987computer}
\bibinfo{author}{\bibfnamefont{Allen} \bibnamefont{M.~P.}}, \bibinfo{author}{\bibfnamefont{Tildesley} \bibnamefont{D.~J.}}, \bibnamefont{et~al.}, \bibinfo{title}{Computer simulation of liquids}, \emph{\bibinfo{journal}{Clarendon-0.12}}  (\bibinfo{year}{1987}).

\bibitem[{\citenamefont{Allen and Tildesley}(2017)}]{allen2017computer}
\bibinfo{author}{\bibfnamefont{Allen} \bibnamefont{M.~P.}} \bibnamefont{and} \bibinfo{author}{\bibfnamefont{Tildesley} \bibnamefont{D.~J.}}, \emph{\bibinfo{title}{Computer simulation of liquids}} (\bibinfo{publisher}{Oxford university press}, \bibinfo{year}{2017}).

\bibitem[{\citenamefont{Guillot}(2002)}]{guillot2002reappraisal}
\bibinfo{author}{\bibfnamefont{Guillot} \bibnamefont{B.}}, \bibinfo{title}{A reappraisal of what we have learnt during three decades of computer simulations on water}, \emph{\bibinfo{journal}{Journal of molecular liquids}} \textbf{\bibinfo{volume}{101}}, \bibinfo{pages}{219} (\bibinfo{year}{2002}).

\bibitem[{\citenamefont{Narten and Levy}(1971)}]{narten1971liquid}
\bibinfo{author}{\bibfnamefont{Narten} \bibnamefont{A.}} \bibnamefont{and} \bibinfo{author}{\bibfnamefont{Levy} \bibnamefont{H.}}, \bibinfo{title}{Liquid water: Molecular correlation functions from x-ray diffraction}, \emph{\bibinfo{journal}{The Journal of Chemical Physics}} \textbf{\bibinfo{volume}{55}}, \bibinfo{pages}{2263} (\bibinfo{year}{1971}).

\bibitem[{\citenamefont{Bandura and Kubicki}(2003)}]{bandura2003derivation}
\bibinfo{author}{\bibfnamefont{Bandura} \bibnamefont{A.}} \bibnamefont{and} \bibinfo{author}{\bibfnamefont{Kubicki} \bibnamefont{J.}}, \bibinfo{title}{Derivation of force field parameters for tio2- h2o systems from ab initio calculations}, \emph{\bibinfo{journal}{The Journal of Physical Chemistry B}} \textbf{\bibinfo{volume}{107}}, \bibinfo{pages}{11072} (\bibinfo{year}{2003}).

\bibitem[{\citenamefont{Rapp{\'e} et~al.}(1992)\citenamefont{Rapp{\'e}, Casewit, Colwell, Goddard~III, and Skiff}}]{rappe1992uff}
\bibinfo{author}{\bibfnamefont{Rapp{\'e}} \bibnamefont{A.~K.}}, \bibinfo{author}{\bibfnamefont{Casewit} \bibnamefont{C.~J.}}, \bibinfo{author}{\bibfnamefont{Colwell} \bibnamefont{K.}}, \bibinfo{author}{\bibfnamefont{Goddard~III} \bibnamefont{W.~A.}}, \bibnamefont{and} \bibinfo{author}{\bibfnamefont{Skiff} \bibnamefont{W.~M.}}, \bibinfo{title}{Uff, a full periodic table force field for molecular mechanics and molecular dynamics simulations}, \emph{\bibinfo{journal}{Journal of the American chemical society}} \textbf{\bibinfo{volume}{114}}, \bibinfo{pages}{10024} (\bibinfo{year}{1992}).

\bibitem[{\citenamefont{Wang et~al.}(2021)\citenamefont{Wang, Huang, Zhang, Wang, Huang, Kuang, and Huang}}]{wang2021diamond}
\bibinfo{author}{\bibfnamefont{Wang} \bibnamefont{J.}}, \bibinfo{author}{\bibfnamefont{Huang} \bibnamefont{X.}}, \bibinfo{author}{\bibfnamefont{Zhang} \bibnamefont{H.}}, \bibinfo{author}{\bibfnamefont{Wang} \bibnamefont{L.}}, \bibinfo{author}{\bibfnamefont{Huang} \bibnamefont{W.}}, \bibinfo{author}{\bibfnamefont{Kuang} \bibnamefont{S.}}, \bibnamefont{and} \bibinfo{author}{\bibfnamefont{Huang} \bibnamefont{F.}}, \bibinfo{title}{Diamond (001)--si (001) and si (001)--ti (0001) interfaces: a density functional theory study}, \emph{\bibinfo{journal}{Journal of Physics and Chemistry of Solids}} \textbf{\bibinfo{volume}{150}}, \bibinfo{pages}{109865} (\bibinfo{year}{2021}).

\bibitem[{\citenamefont{Ikehata et~al.}(2004)\citenamefont{Ikehata, Nagasako, Furuta, Fukumoto, Miwa, and Saito}}]{ikehata2004first}
\bibinfo{author}{\bibfnamefont{Ikehata} \bibnamefont{H.}}, \bibinfo{author}{\bibfnamefont{Nagasako} \bibnamefont{N.}}, \bibinfo{author}{\bibfnamefont{Furuta} \bibnamefont{T.}}, \bibinfo{author}{\bibfnamefont{Fukumoto} \bibnamefont{A.}}, \bibinfo{author}{\bibfnamefont{Miwa} \bibnamefont{K.}}, \bibnamefont{and} \bibinfo{author}{\bibfnamefont{Saito} \bibnamefont{T.}}, \bibinfo{title}{First-principles calculations for development of low elastic modulus ti alloys}, \emph{\bibinfo{journal}{Physical Review B}} \textbf{\bibinfo{volume}{70}}, \bibinfo{pages}{174113} (\bibinfo{year}{2004}).

\bibitem[{\citenamefont{Hua et~al.}(2017)\citenamefont{Hua, Zhang, Kou, Li, Gan, Fundenberger, and Esling}}]{hua2017composite}
\bibinfo{author}{\bibfnamefont{Hua} \bibnamefont{K.}}, \bibinfo{author}{\bibfnamefont{Zhang} \bibnamefont{Y.}}, \bibinfo{author}{\bibfnamefont{Kou} \bibnamefont{H.}}, \bibinfo{author}{\bibfnamefont{Li}~\bibnamefont{J.}}, \bibinfo{author}{\bibfnamefont{Gan} \bibnamefont{W.}}, \bibinfo{author}{\bibfnamefont{Fundenberger} \bibnamefont{J.-J.}}, \bibnamefont{and} \bibinfo{author}{\bibfnamefont{Esling} \bibnamefont{C.}}, \bibinfo{title}{Composite structure of $\alpha$ phase in metastable $\beta$ ti alloys induced by lattice strain during $\beta$ to $\alpha$ phase transformation}, \emph{\bibinfo{journal}{Acta Materialia}} \textbf{\bibinfo{volume}{132}}, \bibinfo{pages}{307} (\bibinfo{year}{2017}).

\bibitem[{\citenamefont{Fisher and Renken}(1964)}]{fisher1964single}
\bibinfo{author}{\bibfnamefont{Fisher} \bibnamefont{E.}} \bibnamefont{and} \bibinfo{author}{\bibfnamefont{Renken} \bibnamefont{C.}}, \bibinfo{title}{Single-crystal elastic moduli and the hcp→ bcc transformation in ti, zr, and hf}, \emph{\bibinfo{journal}{Physical review}} \textbf{\bibinfo{volume}{135}}, \bibinfo{pages}{A482} (\bibinfo{year}{1964}).

\bibitem[{\citenamefont{Song et~al.}(2013)\citenamefont{Song, Li, and Liu}}]{song2013molecular}
\bibinfo{author}{\bibfnamefont{Song} \bibnamefont{F.}}, \bibinfo{author}{\bibfnamefont{Li}~\bibnamefont{B.}}, \bibnamefont{and} \bibinfo{author}{\bibfnamefont{Liu} \bibnamefont{C.}}, \bibinfo{title}{Molecular dynamics simulation of nanosized water droplet spreading in an electric field}, \emph{\bibinfo{journal}{Langmuir}} \textbf{\bibinfo{volume}{29}}, \bibinfo{pages}{4266} (\bibinfo{year}{2013}).

\bibitem[{\citenamefont{Omeje et~al.}(2024)\citenamefont{Omeje, Prada-Rodrigo, Gamet, Di~Maio, Guillemet, Itina, and Sedao}}]{predictwetting}
\bibinfo{author}{\bibfnamefont{Omeje} \bibnamefont{I.~S.}}, \bibinfo{author}{\bibfnamefont{Prada-Rodrigo} \bibnamefont{J.}}, \bibinfo{author}{\bibfnamefont{Gamet} \bibnamefont{E.}}, \bibinfo{author}{\bibfnamefont{Di~Maio} \bibnamefont{Y.}}, \bibinfo{author}{\bibfnamefont{Guillemet} \bibnamefont{R.}}, \bibinfo{author}{\bibfnamefont{Itina} \bibnamefont{T.}}, \bibnamefont{and} \bibinfo{author}{\bibfnamefont{Sedao} \bibnamefont{X.}}, \bibinfo{title}{Prediction of wetting behaviours of femtosecond laser texturized surfaces}, \emph{\bibinfo{journal}{Journal of Laser Micro/Nanoengineering}} \textbf{\bibinfo{volume}{00}}, \bibinfo{pages}{0} (\bibinfo{year}{2024}).

\bibitem[{\citenamefont{Niu and Tang}(2014)}]{niu2014static}
\bibinfo{author}{\bibfnamefont{Niu} \bibnamefont{D.}} \bibnamefont{and} \bibinfo{author}{\bibfnamefont{Tang} \bibnamefont{G.}}, \bibinfo{title}{Static and dynamic behavior of water droplet on solid surfaces with pillar-type nanostructures from molecular dynamics simulation}, \emph{\bibinfo{journal}{International Journal of Heat and Mass Transfer}} \textbf{\bibinfo{volume}{79}}, \bibinfo{pages}{647} (\bibinfo{year}{2014}).

\bibitem[{\citenamefont{Zhao et~al.}(2021)\citenamefont{Zhao, Wang, Pan, Wang, and Zhao}}]{zhao2021molecular}
\bibinfo{author}{\bibfnamefont{Zhao} \bibnamefont{J.}}, \bibinfo{author}{\bibfnamefont{Wang} \bibnamefont{B.}}, \bibinfo{author}{\bibfnamefont{Pan} \bibnamefont{Y.}}, \bibinfo{author}{\bibfnamefont{Wang} \bibnamefont{W.}}, \bibnamefont{and} \bibinfo{author}{\bibfnamefont{Zhao} \bibnamefont{C.}}, \bibinfo{title}{Molecular dynamics simulation on wetting behaviors of n-octane and water droplets on polytetrafluoroethylene surfaces}, \emph{\bibinfo{journal}{Chemical Physics Letters}} \textbf{\bibinfo{volume}{785}}, \bibinfo{pages}{139161} (\bibinfo{year}{2021}).

\bibitem[{\citenamefont{Omeje and Itina}(2022)}]{omeje2022numerical}
\bibinfo{author}{\bibfnamefont{Omeje} \bibnamefont{I.~S.}} \bibnamefont{and} \bibinfo{author}{\bibfnamefont{Itina} \bibnamefont{T.~E.}}, \bibinfo{title}{Numerical study of the wetting dynamics of droplet on laser textured surfaces: Beyond classical wenzel and cassie-baxter model}, \emph{\bibinfo{journal}{Applied Surface Science Advances}} \textbf{\bibinfo{volume}{9}}, \bibinfo{pages}{100250} (\bibinfo{year}{2022}).

\bibitem[{\citenamefont{Frenkel and Smit}(2023)}]{frenkel2023understanding}
\bibinfo{author}{\bibfnamefont{Frenkel} \bibnamefont{D.}} \bibnamefont{and} \bibinfo{author}{\bibfnamefont{Smit} \bibnamefont{B.}}, \emph{\bibinfo{title}{Understanding molecular simulation: from algorithms to applications}} (\bibinfo{publisher}{Elsevier}, \bibinfo{year}{2023}).

\bibitem[{\citenamefont{Yaghoubi and Foroutan}(2018)}]{yaghoubi2018molecular}
\bibinfo{author}{\bibfnamefont{Yaghoubi} \bibnamefont{H.}} \bibnamefont{and} \bibinfo{author}{\bibfnamefont{Foroutan} \bibnamefont{M.}}, \bibinfo{title}{Molecular investigation of the wettability of rough surfaces using molecular dynamics simulation}, \emph{\bibinfo{journal}{Physical Chemistry Chemical Physics}} \textbf{\bibinfo{volume}{20}}, \bibinfo{pages}{22308} (\bibinfo{year}{2018}).

\bibitem[{\citenamefont{Koishi et~al.}(2009)\citenamefont{Koishi, Yasuoka, Fujikawa, Ebisuzaki, and Zeng}}]{koishi2009coexistence}
\bibinfo{author}{\bibfnamefont{Koishi} \bibnamefont{T.}}, \bibinfo{author}{\bibfnamefont{Yasuoka} \bibnamefont{K.}}, \bibinfo{author}{\bibfnamefont{Fujikawa} \bibnamefont{S.}}, \bibinfo{author}{\bibfnamefont{Ebisuzaki} \bibnamefont{T.}}, \bibnamefont{and} \bibinfo{author}{\bibfnamefont{Zeng} \bibnamefont{X.~C.}}, \bibinfo{title}{Coexistence and transition between cassie and wenzel state on pillared hydrophobic surface}, \emph{\bibinfo{journal}{Proceedings of the National Academy of Sciences}} \textbf{\bibinfo{volume}{106}}, \bibinfo{pages}{8435} (\bibinfo{year}{2009}).

\bibitem[{\citenamefont{Plimpton}(1995)}]{plimpton1995fast}
\bibinfo{author}{\bibfnamefont{Plimpton} \bibnamefont{S.}}, \bibinfo{title}{Fast parallel algorithms for short-range molecular dynamics}, \emph{\bibinfo{journal}{Journal of computational physics}} \textbf{\bibinfo{volume}{117}}, \bibinfo{pages}{1} (\bibinfo{year}{1995}).

\bibitem[{\citenamefont{Stukowski}(2009)}]{stukowski2009visualization}
\bibinfo{author}{\bibfnamefont{Stukowski} \bibnamefont{A.}}, \bibinfo{title}{Visualization and analysis of atomistic simulation data with ovito--the open visualization tool}, \emph{\bibinfo{journal}{Modelling and simulation in materials science and engineering}} \textbf{\bibinfo{volume}{18}}, \bibinfo{pages}{015012} (\bibinfo{year}{2009}).

\bibitem[{\citenamefont{Zhang et~al.}(2022)\citenamefont{Zhang, Jiang, Long, Han, Cao, Zhang, Feng, Jia, Sun, Qiu et~al.}}]{zhang2022femtosecond}
\bibinfo{author}{\bibfnamefont{Zhang} \bibnamefont{Y.}}, \bibinfo{author}{\bibfnamefont{Jiang} \bibnamefont{Q.}}, \bibinfo{author}{\bibfnamefont{Long} \bibnamefont{M.}}, \bibinfo{author}{\bibfnamefont{Han} \bibnamefont{R.}}, \bibinfo{author}{\bibfnamefont{Cao} \bibnamefont{K.}}, \bibinfo{author}{\bibfnamefont{Zhang} \bibnamefont{S.}}, \bibinfo{author}{\bibfnamefont{Feng} \bibnamefont{D.}}, \bibinfo{author}{\bibfnamefont{Jia} \bibnamefont{T.}}, \bibinfo{author}{\bibfnamefont{Sun} \bibnamefont{Z.}}, \bibinfo{author}{\bibfnamefont{Qiu} \bibnamefont{J.}}, \bibnamefont{et~al.}, \bibinfo{title}{Femtosecond laser-induced periodic structures: Mechanisms, techniques, and applications}, \emph{\bibinfo{journal}{Opto-Electronic Science}} \textbf{\bibinfo{volume}{1}}, \bibinfo{pages}{220005} (\bibinfo{year}{2022}).

\bibitem[{\citenamefont{Gnilitskyi et~al.}(2017)\citenamefont{Gnilitskyi, Derrien, Levy, Bulgakova, Mocek, and Orazi}}]{gnilitskyi2017high}
\bibinfo{author}{\bibfnamefont{Gnilitskyi} \bibnamefont{I.}}, \bibinfo{author}{\bibfnamefont{Derrien} \bibnamefont{T.~J.-Y.}}, \bibinfo{author}{\bibfnamefont{Levy} \bibnamefont{Y.}}, \bibinfo{author}{\bibfnamefont{Bulgakova} \bibnamefont{N.~M.}}, \bibinfo{author}{\bibfnamefont{Mocek} \bibnamefont{T.}}, \bibnamefont{and} \bibinfo{author}{\bibfnamefont{Orazi} \bibnamefont{L.}}, \bibinfo{title}{High-speed manufacturing of highly regular femtosecond laser-induced periodic surface structures: Physical origin of regularity}, \emph{\bibinfo{journal}{Scientific reports}} \textbf{\bibinfo{volume}{7}}, \bibinfo{pages}{8485} (\bibinfo{year}{2017}).

\bibitem[{\citenamefont{Sampayo et~al.}(2010)\citenamefont{Sampayo, Malijevsk{\`y}, M{\"u}ller, de~Miguel, and Jackson}}]{sampayo2010communications}
\bibinfo{author}{\bibfnamefont{Sampayo} \bibnamefont{J.~G.}}, \bibinfo{author}{\bibfnamefont{Malijevsk{\`y}} \bibnamefont{A.}}, \bibinfo{author}{\bibfnamefont{M{\"u}ller} \bibnamefont{E.~A.}}, \bibinfo{author}{\bibfnamefont{de~Miguel} \bibnamefont{E.}}, \bibnamefont{and} \bibinfo{author}{\bibfnamefont{Jackson} \bibnamefont{G.}}, \bibinfo{title}{Communications: Evidence for the role of fluctuations in the thermodynamics of nanoscale drops and the implications in computations of the surface tension}, \emph{\bibinfo{journal}{The Journal of chemical physics}} \textbf{\bibinfo{volume}{132}} (\bibinfo{year}{2010}).

\bibitem[{\citenamefont{Do~Hong et~al.}(2009)\citenamefont{Do~Hong, Ha, and Balachandar}}]{do2009static}
\bibinfo{author}{\bibfnamefont{Do~Hong} \bibnamefont{S.}}, \bibinfo{author}{\bibfnamefont{Ha}~\bibnamefont{M.~Y.}}, \bibnamefont{and} \bibinfo{author}{\bibfnamefont{Balachandar} \bibnamefont{S.}}, \bibinfo{title}{Static and dynamic contact angles of water droplet on a solid surface using molecular dynamics simulation}, \emph{\bibinfo{journal}{Journal of colloid and interface science}} \textbf{\bibinfo{volume}{339}}, \bibinfo{pages}{187} (\bibinfo{year}{2009}).

\bibitem[{\citenamefont{Gao et~al.}(2018)\citenamefont{Gao, Liao, Liu, and Liu}}]{gao2018nanodroplets}
\bibinfo{author}{\bibfnamefont{Gao} \bibnamefont{S.}}, \bibinfo{author}{\bibfnamefont{Liao} \bibnamefont{Q.}}, \bibinfo{author}{\bibfnamefont{Liu} \bibnamefont{W.}}, \bibnamefont{and} \bibinfo{author}{\bibfnamefont{Liu} \bibnamefont{Z.}}, \bibinfo{title}{Nanodroplets impact on rough surfaces: a simulation and theoretical study}, \emph{\bibinfo{journal}{Langmuir}} \textbf{\bibinfo{volume}{34}}, \bibinfo{pages}{5910} (\bibinfo{year}{2018}).

\bibitem[{\citenamefont{Malijevsk{\`y} and Jackson}(2012)}]{malijevsky2012perspective}
\bibinfo{author}{\bibfnamefont{Malijevsk{\`y}} \bibnamefont{A.}} \bibnamefont{and} \bibinfo{author}{\bibfnamefont{Jackson} \bibnamefont{G.}}, \bibinfo{title}{A perspective on the interfacial properties of nanoscopic liquid drops}, \emph{\bibinfo{journal}{Journal of Physics: Condensed Matter}} \textbf{\bibinfo{volume}{24}}, \bibinfo{pages}{464121} (\bibinfo{year}{2012}).

\bibitem[{\citenamefont{Santiso et~al.}(2013)\citenamefont{Santiso, Herdes, and M{\"u}ller}}]{santiso2013calculation}
\bibinfo{author}{\bibfnamefont{Santiso} \bibnamefont{E.~E.}}, \bibinfo{author}{\bibfnamefont{Herdes} \bibnamefont{C.}}, \bibnamefont{and} \bibinfo{author}{\bibfnamefont{M{\"u}ller} \bibnamefont{E.~A.}}, \bibinfo{title}{On the calculation of solid-fluid contact angles from molecular dynamics}, \emph{\bibinfo{journal}{Entropy}} \textbf{\bibinfo{volume}{15}}, \bibinfo{pages}{3734} (\bibinfo{year}{2013}).

\bibitem[{\citenamefont{{\v{S}}olc et~al.}(2011)\citenamefont{{\v{S}}olc, Gerzabek, Lischka, and Tunega}}]{vsolc2011wettability}
\bibinfo{author}{\bibfnamefont{{\v{S}}olc} \bibnamefont{R.}}, \bibinfo{author}{\bibfnamefont{Gerzabek} \bibnamefont{M.~H.}}, \bibinfo{author}{\bibfnamefont{Lischka} \bibnamefont{H.}}, \bibnamefont{and} \bibinfo{author}{\bibfnamefont{Tunega} \bibnamefont{D.}}, \bibinfo{title}{Wettability of kaolinite (001) surfaces—molecular dynamic study}, \emph{\bibinfo{journal}{Geoderma}} \textbf{\bibinfo{volume}{169}}, \bibinfo{pages}{47} (\bibinfo{year}{2011}).

\bibitem[{\citenamefont{Schneider et~al.}(2012)\citenamefont{Schneider, Rasband, and Eliceiri}}]{schneider2012nih}
\bibinfo{author}{\bibfnamefont{Schneider} \bibnamefont{C.~A.}}, \bibinfo{author}{\bibfnamefont{Rasband} \bibnamefont{W.~S.}}, \bibnamefont{and} \bibinfo{author}{\bibfnamefont{Eliceiri} \bibnamefont{K.~W.}}, \bibinfo{title}{Nih image to imagej: 25 years of image analysis}, \emph{\bibinfo{journal}{Nature methods}} \textbf{\bibinfo{volume}{9}}, \bibinfo{pages}{671} (\bibinfo{year}{2012}).

\bibitem[{\citenamefont{Kim et~al.}(2021)\citenamefont{Kim, Ji, Jang, Alam, Kim, Cho, and Lim}}]{kim2021biological}
\bibinfo{author}{\bibfnamefont{Kim} \bibnamefont{H.-S.}}, \bibinfo{author}{\bibfnamefont{Ji}~\bibnamefont{M.-K.}}, \bibinfo{author}{\bibfnamefont{Jang} \bibnamefont{W.-H.}}, \bibinfo{author}{\bibfnamefont{Alam} \bibnamefont{K.}}, \bibinfo{author}{\bibfnamefont{Kim} \bibnamefont{H.-S.}}, \bibinfo{author}{\bibfnamefont{Cho} \bibnamefont{H.-S.}}, \bibnamefont{and} \bibinfo{author}{\bibfnamefont{Lim} \bibnamefont{H.-P.}}, \bibinfo{title}{Biological effects of the novel mulberry surface characterized by micro/nanopores and plasma-based graphene oxide deposition on titanium}, \emph{\bibinfo{journal}{International Journal of Nanomedicine}}  \bibinfo{pages}{7307--7317} (\bibinfo{year}{2021}).

\bibitem[{\citenamefont{Liu et~al.}(2018)\citenamefont{Liu, Lee, Liu, Hou, Li, Hao, Pan, and Huang}}]{liu2018unique}
\bibinfo{author}{\bibfnamefont{Liu} \bibnamefont{C.-F.}}, \bibinfo{author}{\bibfnamefont{Lee} \bibnamefont{T.-H.}}, \bibinfo{author}{\bibfnamefont{Liu} \bibnamefont{J.-F.}}, \bibinfo{author}{\bibfnamefont{Hou} \bibnamefont{W.-T.}}, \bibinfo{author}{\bibfnamefont{Li}~\bibnamefont{S.-J.}}, \bibinfo{author}{\bibfnamefont{Hao} \bibnamefont{Y.-L.}}, \bibinfo{author}{\bibfnamefont{Pan} \bibnamefont{H.}}, \bibnamefont{and} \bibinfo{author}{\bibfnamefont{Huang} \bibnamefont{H.-H.}}, \bibinfo{title}{A unique hybrid-structured surface produced by rapid electrochemical anodization enhances bio-corrosion resistance and bone cell responses of $\beta$-type ti-24nb-4zr-8sn alloy}, \emph{\bibinfo{journal}{Scientific reports}} \textbf{\bibinfo{volume}{8}}, \bibinfo{pages}{6623} (\bibinfo{year}{2018}).

\bibitem[{\citenamefont{Rad et~al.}(2014)\citenamefont{Rad, Solati-Hashjin, Osman, and Faghihi}}]{rad2014improved}
\bibinfo{author}{\bibfnamefont{Rad} \bibnamefont{A.~T.}}, \bibinfo{author}{\bibfnamefont{Solati-Hashjin} \bibnamefont{M.}}, \bibinfo{author}{\bibfnamefont{Osman} \bibnamefont{N.~A.~A.}}, \bibnamefont{and} \bibinfo{author}{\bibfnamefont{Faghihi} \bibnamefont{S.}}, \bibinfo{title}{Improved bio-physical performance of hydroxyapatite coatings obtained by electrophoretic deposition at dynamic voltage}, \emph{\bibinfo{journal}{Ceramics International}} \textbf{\bibinfo{volume}{40}}, \bibinfo{pages}{12681} (\bibinfo{year}{2014}).

\bibitem[{\citenamefont{Goodarzi et~al.}(2016)\citenamefont{Goodarzi, Moztarzadeh, Nezafati, and Omidvar}}]{goodarzi2016titanium}
\bibinfo{author}{\bibfnamefont{Goodarzi} \bibnamefont{S.}}, \bibinfo{author}{\bibfnamefont{Moztarzadeh} \bibnamefont{F.}}, \bibinfo{author}{\bibfnamefont{Nezafati} \bibnamefont{N.}}, \bibnamefont{and} \bibinfo{author}{\bibfnamefont{Omidvar} \bibnamefont{H.}}, \bibinfo{title}{Titanium dioxide nanotube arrays: A novel approach into periodontal tissue regeneration on the surface of titanium implants}, \emph{\bibinfo{journal}{Advanced Materials Letters}} \textbf{\bibinfo{volume}{7}}, \bibinfo{pages}{209} (\bibinfo{year}{2016}).

\bibitem[{\citenamefont{Alen et~al.}(2016)\citenamefont{Alen, Farhat, and Rahman}}]{alen2016analytical}
\bibinfo{author}{\bibfnamefont{Alen} \bibnamefont{S.~K.}}, \bibinfo{author}{\bibfnamefont{Farhat} \bibnamefont{N.}}, \bibnamefont{and} \bibinfo{author}{\bibfnamefont{Rahman} \bibnamefont{M.~A.}}, in \emph{\bibinfo{booktitle}{AIP Conference Proceedings}} (\bibinfo{organization}{AIP Publishing}, \bibinfo{year}{2016}), vol. \bibinfo{volume}{1754}.

\bibitem[{\citenamefont{Hay et~al.}(2008)\citenamefont{Hay, Dragila, and Liburdy}}]{hay2008theoretical}
\bibinfo{author}{\bibfnamefont{Hay} \bibnamefont{K.}}, \bibinfo{author}{\bibfnamefont{Dragila} \bibnamefont{M.}}, \bibnamefont{and} \bibinfo{author}{\bibfnamefont{Liburdy} \bibnamefont{J.}}, \bibinfo{title}{Theoretical model for the wetting of a rough surface}, \emph{\bibinfo{journal}{Journal of colloid and interface science}} \textbf{\bibinfo{volume}{325}}, \bibinfo{pages}{472} (\bibinfo{year}{2008}).

\bibitem[{\citenamefont{Zhang et~al.}(2016)\citenamefont{Zhang, Matin, Ha, and Jang}}]{zhang2016molecular}
\bibinfo{author}{\bibfnamefont{Zhang} \bibnamefont{Z.}}, \bibinfo{author}{\bibfnamefont{Matin} \bibnamefont{M.~A.}}, \bibinfo{author}{\bibfnamefont{Ha}~\bibnamefont{M.~Y.}}, \bibnamefont{and} \bibinfo{author}{\bibfnamefont{Jang} \bibnamefont{J.}}, \bibinfo{title}{Molecular dynamics study of the hydrophilic-to-hydrophobic switching in the wettability of a gold surface corrugated with spherical cavities}, \emph{\bibinfo{journal}{Langmuir}} \textbf{\bibinfo{volume}{32}}, \bibinfo{pages}{9658} (\bibinfo{year}{2016}).

\end{thebibliography}

\end{document}